\documentclass{article}

\usepackage{PRIMEarxiv}

\usepackage[utf8]{inputenc} 
\usepackage[T1]{fontenc}    
\usepackage{hyperref}       
\usepackage{url}            
\usepackage{booktabs}       
\usepackage{amsfonts}       
\usepackage{nicefrac}       
\usepackage{microtype}      
\usepackage{lipsum}
\usepackage{fancyhdr}       
\usepackage{graphicx}       
\graphicspath{{media/}}     

\usepackage{amsmath}
\usepackage{multirow}
\usepackage[table,xcdraw]{xcolor}

\title{Benchmarking MedMNIST dataset on real quantum hardware}

\author{
  Gurinder Singh, Hongni Jin, and Kenneth M. Merz, Jr.  \\
  Center for Computational Life Sciences, Lerner Research Institute \\ Cleveland Clinic, Cleveland, Ohio 44106, United States \\
  \texttt{\{singhg12, jinh2, merzk\}@ccf.org}
}

\begin{document}
\maketitle

\begin{abstract}
Quantum machine learning (QML) has emerged as a promising domain to leverage the computational capabilities of quantum systems to solve complex classification tasks. In this work, we present the first comprehensive QML study by benchmarking the MedMNIST-a diverse collection of medical imaging datasets on a 127-qubit real IBM quantum hardware, to evaluate the feasibility and performance of quantum models (without any classical neural networks) in practical applications. This study explores recent advancements in quantum computing such as device-aware quantum circuits, error suppression, and mitigation for medical image classification. Our methodology is comprised of three stages: preprocessing, generation of noise-resilient and hardware-efficient quantum circuits, optimizing/training of quantum circuits on classical hardware, and inference on real IBM quantum hardware. Firstly, we process all input images in the preprocessing stage to reduce the spatial dimension due to quantum hardware limitations. We generate hardware-efficient quantum circuits using backend properties expressible to learn complex patterns for medical image classification. After classical optimization of QML models, we perform inference on real quantum hardware. We also incorporate advanced error suppression and mitigation techniques in our QML workflow, including dynamical decoupling (DD), gate twirling, and matrix-free measurement mitigation (M3) to mitigate the effects of noise and improve classification performance. The experimental results showcase the potential of quantum computing for medical imaging and establish a benchmark for future advancements in QML applied to healthcare.

\end{abstract}

\keywords{Quantum machine learning \and Medical image classification \and Device-aware quantum circuit \and Error mitigation}

\section{Introduction}

Biomedical image analysis plays a critical role in advancing human healthcare. Recently, machine learning (ML) has demonstrated remarkable potential in this domain, excelling in tasks such as medical image classification \cite{Yang_2023,kumar2024medical,yue2024medmambavisionmambamedical}, segmentation \cite{Ma_2024,zhang2024segmentmodelmedicalimage}, and reconstruction \cite{webber2024diffusion,khader2023medicaldiffusiondenoisingdiffusion}. Moreover, QML, an emerging paradigm of ML implemented on quantum computers, is anticipated to offer quantum advantages over classical ML. These advantages include enhanced speed-ups, improved model expressivity, efficient exploration of high-dimensional feature spaces, and the ability to address certain classically intractable problems \cite{Liu_2021,Abbas_2021,jin2024integratingmachinelearningquantum,yamasaki2023advantagequantummachinelearning,xiao2023practical}. QML capitalizes on the unique properties of quantum mechanics, such as superposition, entanglement, and interference, to unravel complex relationships between inputs and labels. In the current noisy intermediate-scale quantum (NISQ) era, QML faces notable challenges. Limited access to large-scale quantum systems constrains the ability to encode high-dimensional data into quantum states. Hardware noise and decoherence further impact accuracy and restrict circuit depth, thereby limiting the expressivity of QML models. Rapid advancements in quantum hardware, including improved quantum error correction techniques \cite{acharya2024quantumerrorcorrectionsurface}, are expected to overcome these limitations and enable the practical deployment of QML for real-world applications in the near future.

Despite these obstacles, QML research has witnessed significant growth \cite{Grant_2018,farhi2018classificationquantumneuralnetworks,Havl_ek_2019,kerenidis2019quantumalgorithmsdeepconvolutional,Cong_2019,allcock2019quantumalgorithmsfeedforwardneural,hernández2020imageclassificationquantummachine,kiani2020quantummedicalimagingalgorithms,Nakaji_2021,schuld2021effect,schuld2021quantum,kollias2023quantum,cherrat2024quantum,fan2023hybrid,wang2024shallow}. Generally, QML approaches are based on parameterized quantum circuits (PQCs) \cite{Benedetti_2019} consisting of fixed gates (such as SWAP and CNOT) and single-qubit (such as Pauli rotation) gates with trainable parameters. These PQCs are comprised of three stages: data encoding, a variational ansatz circuit with trainable parameters, and measurement to classical bits. The data encoding stage encodes the classical input data to quantum states. The parameters of the ansatz circuit are optimized using classical gradient descent algorithms, typically relying on the parameter-shift rule \cite{mitarai2018quantum}, enabling the circuit to solve machine learning tasks.

Over the last few years, researchers have explored QML algorithms for medical imaging in various studies including \cite{cherrat2024quantum,Adhikary_2020,Schuld_2020,acar2021covid,moradi2022clinical,houssein2022hybrid,toledo2022grading,azevedo2022quantum,parisi2022quantum,landman2022quantum} to harness potential quantum advantages over classical machine learning approaches. These studies address a variety of medical imaging modalities, including MRI, CT, X-ray, OCT, and ultrasound, targeting tasks such as clinical data classification and detection. The QML algorithms investigated in these works include quantum support vector machines (SVM) \cite{moradi2022clinical}, hybrid quantum-classical neural networks \cite{cherrat2024quantum,houssein2022hybrid,toledo2022grading,parisi2022quantum,landman2022quantum}, quantum transfer learning \cite{acar2021covid,azevedo2022quantum}, and variational quantum circuits \cite{Adhikary_2020,Schuld_2020}. To the best of our knowledge, few studies such as \cite{cherrat2024quantum,acar2021covid,moradi2022clinical,azevedo2022quantum,landman2022quantum} have used real quantum hardware for evaluation purposes but on a limited scale. Cherrat et al. proposed a quantum vision transformer (QViT) in \cite{cherrat2024quantum} based on shallow PQCs capable of performing complex image classification tasks. QViT introduces trainable quantum orthogonal layers adaptable to varying levels of quantum hardware connectivity. The authors have evaluated QViT on open-source MedMNIST (a collection of 12 biomedical imaging datasets) \cite{Yang_2023} using a simulator and real hardware results are provided for only one dataset (RetinaMNIST) out of 12 total. In \cite{acar2021covid}, Acar and Yilmaz designed a hybrid classical-quantum transfer learning based QML model for COVID-19 detection using CT images. The authors claimed to achieve a classification accuracy of 94-100\% on quantum devices with limited qubit capacity compared to 90\% on classical systems using a small dataset. This highlights the potential of quantum computers for efficient and accurate medical image classification, particularly with limited data. Moradi et al. \cite{moradi2022clinical} designed two QML methods i.e., quantum distance classifier and quantum-kernel SVM for clinical data classification on quantum hardware. The findings observed in this study also highlight the promise of QML in clinical data analysis, especially for tasks constrained by dataset size and feature complexity. Azevedo et al. \cite{azevedo2022quantum} employed a hybrid quantum-classical neural network for mammogram classification into malignant and benign categories. The authors compared the results obtained from the quantum device and simulator, highlighting the potential of QML in medical diagnostics while emphasizing the need for further validation. Similarly, the study conducted in \cite{landman2022quantum} by Landman et al. introduces two quantum neural networks: quantum-assisted neural networks, which leverage the quantum computer to estimate inner products during the training and inference of classical neural networks, and quantum orthogonal neural networks, built on quantum pyramidal circuits for orthogonal matrix multiplication. This study also considered only two datasets (PneumoniaMNIST and RetinaMNIST) out of 12 datasets from MedMNIST for evaluation purposes on real hardware. It is also worth noting that most of these QML studies are based on hybrid quantum-classical neural networks that combine quantum circuits with classical neural networks, leaving the potential of purely quantum approaches underexplored. Also, the individual contribution of both quantum and classical parts is unclear. While some studies claim potential quantum advantages, these are often observed only in restricted settings (e.g., small datasets) with limited evidence of their generalization across different image classification domains or modalities. Demonstrating clear, reproducible quantum advantages over classical machine learning models remains a significant challenge. 

To address above mentioned limitations, we conducted a comprehensive QML study for medical image classification on a 127 qubits IBM quantum hardware. This research study aims to explore the different aspects of QML for image classification including efficient encoding of high-dimensional image data to quantum states, generation of device-aware quantum circuits having optimal expressivity for different image datasets, advantages of error suppression and mitigation techniques. By exclusively employing quantum methods without classical neural networks, our work addresses critical gaps in the current QML literature, offering insights into the practical applicability and performance of quantum-only models for medical image classification tasks. This effort not only highlights the promise of QML in addressing real-world problems but also sets a foundational benchmark for future explorations in this domain. The main contributions of our research study are as follows: 

\begin{itemize}
\item We present the first comprehensive QML study by benchmarking the MedMNIST dataset on IBM Cleveland.
\item This research leverages and evaluates the recent quantum computing capabilities such as generation of device-aware quantum circuits and error suppression/mitigation techniques for medical image classification.
\item We also conducted an ablation study on different error suppression and mitigation techniques to provide an intuitive understanding.
\item The experimental results provide foundational insights into the performance and scalability of quantum models for medical image classification, establishing a benchmark for future QML research for healthcare applications.
\end{itemize}

In the remaining sections of this research study, we first discussed our QML methodology in detail in Section 2. Then, we provide the experimental results with detailed discussion in Section 3. Lastly, concluding remarks are provided in Section 4.   

\section{Methodology}

\begin{figure}[!ht]
    \centering
    \includegraphics[width=\linewidth]{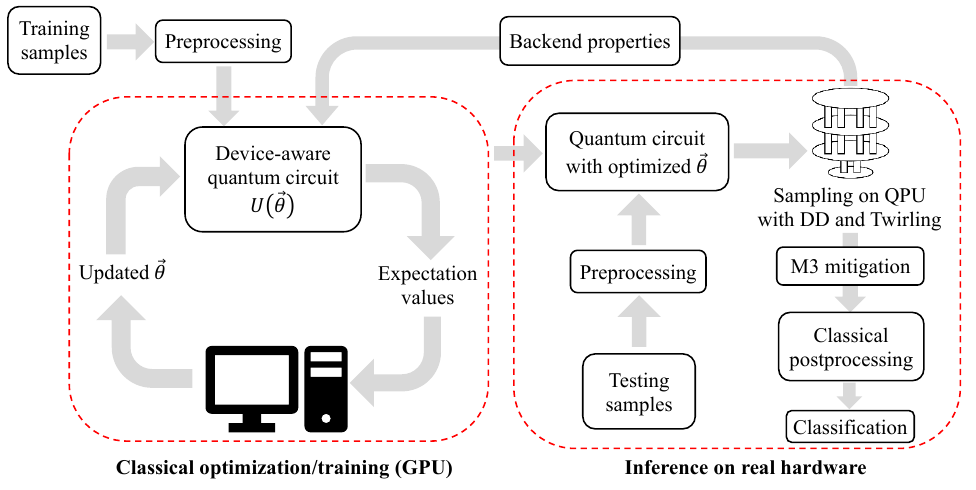}
    \caption{QML framework for medical image classification.}
    \label{figure1}
\end{figure}

In this section, we discuss our QML methodology for medical image classification, as illustrated in Fig. \ref{figure1}. Our approach is structured into four key stages: data preprocessing, generation of device-aware quantum circuits, training of the quantum circuits on classical hardware using a training dataset, and inference of the trained quantum model on real IBM quantum hardware using a testing dataset. It is important to emphasize that, due to the limitations of current NISQ quantum devices, training quantum models with large datasets directly on hardware remains infeasible. To address this challenge, our QML workflow adopts a hybrid strategy, where the training phase is executed on classical hardware, leveraging their computational power, while the inference stage is performed on real quantum hardware to validate the model's performance. The following subsections provide a detailed explanation of each stage of our QML methodology.

\subsection{Data preprocessing}

The preprocessing stage focuses on reducing the feature dimensionality of input images, a necessary step in our QML workflow due to the constraints of current quantum hardware. Specifically, we employ angle encoding to map classical data to quantum states. The angle encoding method is limited by the hardware's inability to handle high-dimensional feature spaces in the current NISQ era. Therefore, it is very important to reduce the spatial dimension of the images effectively. We used average pooling to downscale the input image dimension from 224$\times$224 to 7$\times$7 and 8$\times$8 for the considered MedMNIST datasets, thereby providing 49 and 64 features, respectively. This indicates that performing angle encoding for 49 and 64 features requires the incorporation of 49 and 64 Pauli rotation gates, respectively, within the quantum circuit. The quantum circuit representational capacity should be adjusted accordingly to effectively capture the patterns in the input features. Post-feature reduction, the pixel values are normalized to the range $[0, \pi]$, aligning with the requirements of angle encoding for efficient quantum state preparation. It is important to note that this preprocessing stage introduces information loss compared to classical machine learning models, which can process entire images as input without dimensionality reduction. In a QML workflow, this limitation arises due to the constraints of current quantum hardware, which restrict the number of features that can be encoded into quantum states. Consequently, while classical models can leverage the full spatial resolution of input images, quantum models operate with a reduced feature set, potentially impacting the overall representational capacity. This trade-off underscores the need for effective feature-reduction techniques that preserve essential information while ensuring compatibility with the limitations of quantum devices.

\subsection{Efficient and noise-resilient quantum circuits for QML}

The second crucial step of our QML framework involves designing a noise-robust and efficient quantum circuit that is expressively powerful enough to learn meaningful patterns from the input data for the medical imaging classification tasks. In the current NISQ era, developing device-aware quantum circuits tailored to the target backend properties, such as qubit connectivity and gate fidelities, has emerged as a key strategy to mitigate the impact of hardware noise. However, designing such circuits remains a significant challenge due to the lack of well-defined guidelines for quantum circuit architecture in QML workflows. Additionally, the optimal design of a quantum circuit is highly dataset-dependent, as varying datasets require different levels of expressivity to effectively capture their underlying patterns.

To address these challenges, we incorporated the Élivágar framework \cite{anagolum2024elivagarefficientquantumcircuit} into our QML workflow to generate hardware-aware quantum circuits tailored to our medical image classification task. This framework allowed us to create noise-resilient and computationally efficient quantum circuits that align with the physical constraints of the target quantum hardware while maximizing model performance. The first step of the Élivágar framework is to generate candidate quantum circuits using information from the target quantum device. We generated 250 quantum circuits in all the MedMNIST dataset experiments. These quantum circuits include three key components: (1) data encoding gates, (2) a variational or ansatz circuit, and (3) measurement operations that map quantum states to classical bits. The data encoding layer leverages parameterized rotation gates to embed classical input data into quantum states. The ansatz, or variational circuit, contains trainable Pauli rotation gates interspersed with CNOT gates to ensure qubit entanglement. The parameters of these gates ($\theta$) are optimized during training to minimize the loss function. Input data from the preprocessing stage is encoded using some rotation gates equal to the feature count of each input sample. Then, Clifford Noise Resilience (CNR) \cite{anagolum2024elivagarefficientquantumcircuit} is computed to assess the robustness of these candidate quantum circuits against hardware noise. This process involves the generation and execution of multiple Clifford replicas of the circuit (\( C \)) denoted by \( C_R(i) \), followed by a fidelity assessment between the noisy (\( P_{noisy}\)) and noiseless (\( P_{noiseless} \)) outputs. To achieve this, the Total Variation Distance (TVD) is first evaluated between noiseless and noise outputs as defined in the Eq. (\ref{equation1}) and the circuit fidelity $(F)$ is computed by $F = 1 - TVD$. 

\begin{equation}
\label{equation1}
TVD = \frac{1}{2} \sum_{i \in [2^{n_q}]} \left| P_{noiseless}(|i\rangle) - P_{noisy}(|i\rangle) \right|
\end{equation}

where $n_q$ denotes the number of qubits in the circuit. The CNR of a circuit is then computed as the average fidelity of the \( M \) Clifford replicas defined as:

\begin{equation}
CNR(C) = \frac{1}{M} \sum_{i=1}^{M} F( P_{noiseless}^{C_R(i)}, P_{noisy}^{C_R(i)})
\end{equation}

The low-fidelity candidate circuits with CNR values lower than the threshold value (0.7) are excluded from further evaluation. While this process is typically conducted on a quantum device, it is time-intensive due to the need to run thousands of circuits, often facing significant queuing delays. To streamline the process, we performed CNR calculations using IBM Cleveland's hardware noise model on a simulator (with the number of shots = 10000 and \( M \) = 32), enabling efficient evaluation of circuit noise resilience while saving considerable time. Afterward, the performance of selected quantum circuits is assessed using Representation Capacity (RepCap) \cite{anagolum2024elivagarefficientquantumcircuit} defined as:

\begin{equation}
RepCap(C) = 1 - \frac{\|R_C - R_{\text{ref}}\|_2^2}{2 n_c d_c^2}
\end{equation}

where \( d_c \) represents the number of samples selected per class, \( n_c \) is the number of classes in the training dataset, and \( \|\cdot\|_2 \) denotes the Frobenius norm. We used \( d_c \) = 16 for all the MedMNIST dataset experiments. The matrix \( R_C \) encodes the pairwise similarities between the quantum representations of data samples, while \( R_{\text{ref}} \) is a reference matrix capturing the ideal behavior of the circuit, where \( R_{\text{ref}}(i,j) = 1 \) if two samples belong to the same class and \( 0 \) otherwise. A higher RepCap value suggests that the circuit has a greater capacity to distinguish between different classes in the dataset. Finally, a composite scoring metric (\( F_{score} \)) is computed based on CNR and RepCap to effectively select the most suitable quantum circuit defined as:

\begin{equation}
F_{score}(C) = CNR(C)^{\alpha_{\text{CNR}}} \times RepCap(C)
\end{equation}

where, \( \alpha_{\text{CNR}} \) is a tunable hyperparameter that controls the relative weight of noise resilience in the scoring process and we set \( \alpha_{\text{CNR}} = 0.5 \) in our MedMNIST experiments.

To ensure adaptability and effectiveness, we designed multiple quantum circuits with varying levels of expressivity by systematically adjusting the number of trainable parameters and the depth of the ansatz. These circuits were carefully tailored to meet the unique requirements of the MedMNIST datasets. Specifically, we generated quantum circuits with different configurations, incorporating varying numbers of trainable parameters—such as 60, 80, 100, and 120 for all the MedMNIST datasets under consideration. This approach allowed us to explore the impact of circuit expressivity on the performance of our quantum models and ensured optimal alignment with the specific characteristics of each dataset. Importantly, to thoroughly investigate the potential of quantum circuits in medical imaging classification, our QML framework exclusively employs quantum models without incorporating any classical neural network components. This approach highlights the intrinsic capabilities of quantum circuits and paves the way for future advancements in leveraging QML for medical imaging tasks.

\subsection{Training quantum model on classical hardware}

The training phase of our QML framework for medical image classification is executed on classical hardware as illustrated in Fig. \ref{figure1}. This approach circumvents the current limitations of quantum hardware in the NISQ era, such as limited qubits, high noise levels, and decoherence effects by performing the computationally expensive training process on classical GPUs even for large-scale datasets. Our QML workflow integrates classical optimization techniques with noise-resilient, hardware-efficient quantum circuits tailored for medical imaging tasks. The training pipeline includes iterative optimization using a gradient-based method and loss evaluation. The variational quantum circuit parameters $(\theta)$ are iteratively updated within a classical optimization loop to minimize a task-specific loss function. This iterative process continues until convergence is achieved or the maximum number of training epochs is reached, ensuring the parameters are optimized for the target classification task.

We conducted the training on noiseless simulators using the TorchQuantum library \cite{hanruiwang2022quantumnas}, which provides an ideal environment for parameter optimization without interference from hardware-induced noise. This approach significantly reduces computational overhead compared to direct hardware training and ensures consistent gradient computation. During each training iteration, the preprocessed input data is mapped to quantum states via encoding gates in the quantum circuit, which is then executed on the simulator. Expectation values obtained from circuit measurements are used to compute the loss function. Gradients of the loss function are calculated using backpropagation. Afterward, these gradients guide precise updates to the trainable parameters $(\theta)$. In our implementation, all the quantum circuits were trained for 200 epochs using the Adam optimizer with a learning rate of 0.01 and a batch size of 128, ensuring convergence to optimal parameter configurations. The MedMNIST contains datasets having a different number of classes. During training, we measure one qubit and two qubits for two class and four class datasets, respectively and a mean square loss function is used to calculate the loss. For the remaining multi-class datasets, we measure the number of qubits equal to the number of classes in the dataset, and the cross-entropy loss function is used to calculate the loss. To ensure compatibility with MedMNIST datasets, we trained circuits with varying numbers of trainable parameters, such as 60, 80, 100, and 120 for each dataset of MedMNIST to account for differences in dataset complexity. The best-performing models are saved along with their parameters, losses, and accuracies for later inference on real quantum hardware. We also performed inference of these trained quantum models on a simulator for all considered MedMNIST datasets. All quantum models are trained on an A100 Nvidia GPU with 43 GB RAM.

\subsection{Inference on real IBM Cleveland hardware}

The trained quantum models demonstrating the best classification performance on the simulator were selected for validation on the IBM Cleveland quantum hardware. During the inference stage of our QML framework, the trained quantum models with optimized parameters were utilized for evaluating the test samples. For each testing sample, the corresponding quantum circuit is dynamically constructed and transpiled at optimization level 3 to minimize circuit depth and gate count while ensuring compatibility with the hardware topology. Moreover, qubit connectivity was carefully configured to align with the hardware's topology, minimizing the need for SWAP gates, which can introduce unnecessary noise and overhead. To further enhance the quantum model performance, we employed error suppression techniques, i.e. dynamical decoupling(DD) and gate twirling (Twir) during the sampling process on real IBM hardware. The DD is used to suppress the decoherence effects introduced due to the unwanted interaction between the qubits and the environment. Gate twirling applies random single-qubit gates around a target gate to average out coherent errors. The sampling on real hardware is performed using Qiskit SamplerV2 in Session execution mode. Moreover, we also integrated M3 error mitigation \cite{Nation_2021} in our QML workflow. M3 mitigation is a scalable quantum error correction approach that operates within the noisy bitstring subspace, correcting correlated and uncorrelated errors using a memory-efficient, matrix-free iterative solver. The integration of these error suppression and mitigation strategies enables quantum models to achieve optimal inference performance on real quantum hardware. Lastly, classical post-processing is applied on the mitigated bitstrings to obtain the predicted classification labels. To achieve this, the mitigated bitstrings are processed to compute qubit probabilities defined as:

\begin{equation}
P(q, m) = \frac{N_{q,m}}{N_{\text{shots}}}, \quad m \in \{0, 1\}
\end{equation}

where $P(q,m)$ is the probability of measuring state $m$ on qubit $q$, $N_{q,m}$ is the number of times the $q$-th qubit is measured in state $\lvert m \rangle$ during the experiment. The measurement outcome $m=0$ corresponds to the state $\lvert 0 \rangle$ and $m=1$ corresponds to the state $\lvert 1 \rangle$. $N_{\text{shots}}$ represents the total number of measurement shots and we used $N_{\text{shots}}$ = 32000 for all the MedMNIST datasets. These probabilities provide the final expectation values for classification computed as:

\begin{equation}
\langle Z_q \rangle = P(q, 0) - P(q, 1)
\end{equation}

where $\langle Z_q \rangle$ denotes the expectation value of the Pauli-Z operator for the $q$-th qubit, calculated as the difference between the probabilities of measuring 0 and 1. This expectation value vector obtained for each test sample is processed to obtain the predicted class labels. The results from all test samples were aggregated to compute the performance metrics, providing insights into the performance of the quantum circuits on real quantum hardware. By conducting inference on real quantum device, we demonstrated the feasibility of deploying noise-resilient and device-aware quantum circuits for practical medical image classification tasks.

\section{Experimental results}

\subsection{Dataset and evaluation metrics details}

\begin{figure}[!ht]
    \centering
    \includegraphics[width=0.9\linewidth]{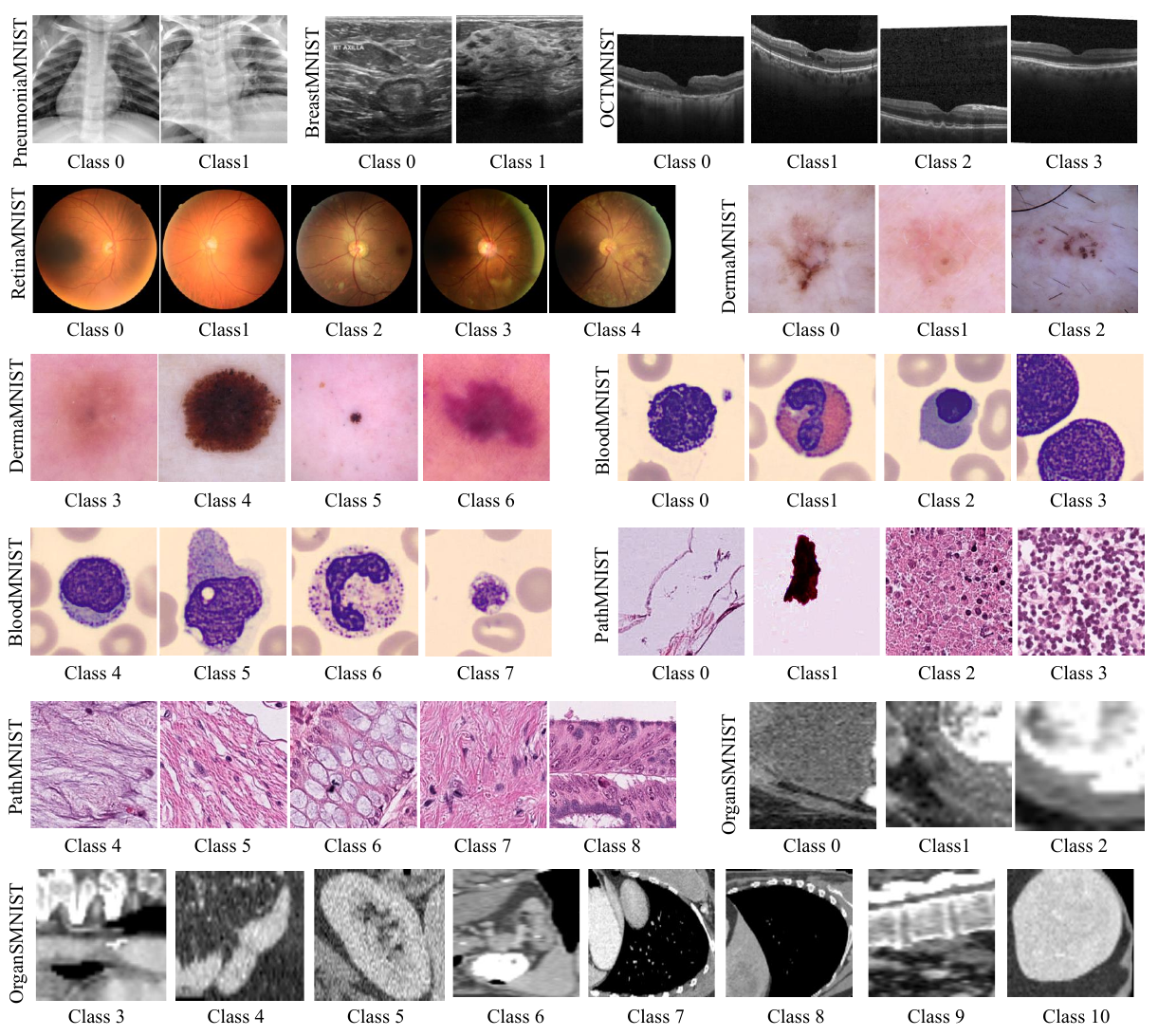}
    \caption{MedMNIST datasets for 2D biomedical image classification.}
    \label{figure2}
\end{figure}

In this subsection, we provide the details of the MedMNIST datasets \cite{Yang_2023} used in this benchmark study and performance metrics used to evaluate the QML workflow. The MedMNIST contains 12 two-dimensional biomedical imaging datasets of different medical modalities and complexity levels. These 12 datasets include 2-class as well as multi-class datasets including 4-class, 5-class, 7-class, 8-class, 9-class, and 11-class. For this study, we considered 8 datasets out of the 12 by making sure that at least one dataset from each class category was considered as shown in Table \ref{table1}. The different medical modalities considered in these 8 datasets are illustrated in Fig. \ref{figure2}. We considered two widely used classification metrics i.e., accuracy (ACC) and area under the curve (AUC) to measure the performance of QML models. It is important to note that ACC provides a straightforward interpretation of model performance but it is highly sensitive to class imbalance, often leading to biased evaluations. On the contrary, AUC offers a more comprehensive evaluation of the classifier's separability and is inherently robust to class imbalance. This makes AUC particularly suitable for datasets where class distributions are skewed. It is clear from Table \ref{table1} that the MedMNIST datasets are quite imbalanced. Therefore, we prioritized AUC over ACC to select a model for real hardware execution. By focusing on AUC, we ensured that the chosen quantum models were able to effectively distinguish between classes across various thresholds, providing a fairer and more reliable assessment of performance. 

\begin{table}[!ht]
\centering
\caption{Data summary of MedMNIST-v2 datasets used in our QML workflow.}
\label{table1}
\scalebox{0.782}{
\begin{tabular}{|c|c|c|c|c|}
\hline
\textbf{MedMNIST2D} & \textbf{Data Modality} & \textbf{Classes} & \textbf{Training / Test} & \textbf{Sample distribution}                                                                                                                                                    \\ \hline
PneumoniaMNIST      & Chest X-Ray            & 2                & 4,708 / 624              & Train: 1214, 3494; Test: 234, 390                                                                                                                                               \\ \hline
BreastMNIST         & Breast Ultrasound      & 2                & 546 / 156                & Train: 147, 399; Test: 42, 114                                                                                                                                                  \\ \hline
OCTMNIST            & Retinal OCT            & 4                & 97,477 / 1,000           & Train: 33484, 10213, 7754, 46026; Test: 250, 250, 250, 250                                                                                                                      \\ \hline
RetinaMNIST         & Fundus Camera          & 5                & 1,080 / 400              & Train: 486, 128, 206, 194, 66; Test: 174, 46, 92, 68, 20                                                                                                                        \\ \hline
DermaMNIST          & Dermatoscope           & 7                & 7,007 / 2,005            & \begin{tabular}[c]{@{}c@{}}Train: 228, 359, 769, 80, 779, 4693, 99\\ Test: 66, 103, 220, 23, 223, 1341, 29\end{tabular}                                                         \\ \hline
BloodMNIST          & Blood Cell Microscope  & 8                & 11,959 / 3,421           & \begin{tabular}[c]{@{}c@{}}Train: 852, 2181, 1085, 2026, 849, 993, 2330, 1643\\      Test: 244, 624, 311, 579, 243, 284, 666, 470\end{tabular}                                  \\ \hline
PathMNIST           & Colon Pathology        & 9                & 89,996 / 7,180           & \begin{tabular}[c]{@{}c@{}}Train: 9366, 9509, 10360, 10401, 8006, 12182, 7886, 9401, 12885\\      Test: 1338, 847, 339, 634, 1035, 592, 741, 421, 1233\end{tabular}             \\ \hline
OrganSMNIST         & Abdominal CT           & 11               & 13,932 / 8,827           & \begin{tabular}[c]{@{}c@{}}Train: 1148, 630, 614, 721, 1132, 1119, 3464, 741, 803, 2004, 1556\\      Test: 811, 439, 445, 510, 704, 693, 2078, 397, 439, 1343, 968\end{tabular} \\ \hline
\end{tabular}}
\end{table}

The classical post-processing of M3-mitigated bitstrings in the inference stage provides an expectation values vector for all test samples. These expectation values are used to calculate the ACC and AUC metrics. For binary datasets (where we measure one qubit), the accuracy is calculated by measuring the proportion of predictions that deviate from the true labels by less than a threshold of 1. It evaluates how many predictions are close enough to the actual labels, providing an overall measure of correctness for the model's outputs in binary classification tasks. The binary AUC is computed by assessing the separability of the two classes on the basis of the predicted expectation values. We generate signed prediction labels to align with the true labels and calculate the AUC score using the AUC score function \cite{scikit-learn} from scikit-learn. For the multi-class datasets (where we are measuring more than 2 qubits), the model's predicted class is determined by taking the index of the maximum value along each row of the expectation values vector. This is compared against the true labels to compute the classification accuracy as:

\begin{equation}
\text{ACC} = \frac{\text{Number of correct predictions}}{\text{Total number of samples}}
\end{equation}

To calculate the AUC metric for multi-class datasets, we first applied the softmax activation function to convert the raw expectation values into class probabilities. This ensures that the outputs for each test sample sum to 1 and can be interpreted as probabilities for each class. Then, we use the AUC function \cite{scikit-learn} from scikit-learn to calculate the AUC. 

\begin{table}[!ht]
\centering
\caption{Performance analysis of QML framework on simulator based on classification accuracy (ACC) and area under the curve (AUC) metrics under different configurations of quantum models.}
\label{table2}
\begin{tabular}{|c|c|c|c|c|c|}
\hline
\cellcolor[HTML]{FFFFFF}\textbf{Dataset} & \cellcolor[HTML]{FFFFFF}\textbf{\begin{tabular}[c]{@{}c@{}}Number of \\ qubits\end{tabular}} & \textbf{\begin{tabular}[c]{@{}c@{}}Number of \\ embedding gates\end{tabular}} & \cellcolor[HTML]{FFFFFF}\textbf{\begin{tabular}[c]{@{}c@{}}Number of   \\ trainable parameters\end{tabular}} & \cellcolor[HTML]{FFFFFF}\textbf{\begin{tabular}[c]{@{}c@{}}Noiseless \\ ACC\end{tabular}} & \cellcolor[HTML]{FFFFFF}\textbf{\begin{tabular}[c]{@{}c@{}}Noiseless \\      AUC\end{tabular}} \\ \hline
                                         &                                                                                              &                                                                               & 60                                                                                                           & 0.8478                                                                                    & 0.8098                                                                                         \\ \cline{4-6} 
                                         &                                                                                              &                                                                               & 80                                                                                                           & 0.8462                                                                                    & 0.8137                                                                                         \\ \cline{4-6} 
                                         &                                                                                              &                                                                               & 100                                                                                                          & 0.8462                                                                                    & 0.8068                                                                                         \\ \cline{4-6} 
                                         &                                                                                              & \multirow{-4}{*}{49}                                                          & 120                                                                                                          & \textbf{0.8526}                                                                           & \textbf{0.8197}                                                                                \\ \cline{3-6} 
                                         &                                                                                              &                                                                               & 60                                                                                                           & 0.8253                                                                                    & 0.785                                                                                          \\ \cline{4-6} 
                                         &                                                                                              &                                                                               & 80                                                                                                           & 0.8462                                                                                    & 0.8128                                                                                         \\ \cline{4-6} 
                                         &                                                                                              &                                                                               & 100                                                                                                          & \textbf{0.8526}                                                                                    & 0.812                                                                                          \\ \cline{4-6} 
\multirow{-8}{*}{PneumoniaMNIST}         & \multirow{-8}{*}{4}                                                                          & \multirow{-4}{*}{64}                                                          & 120                                                                                                          & 0.8478                                                                                    & 0.8132                                                                                         \\ \hline
                                         &                                                                                              &                                                                               & 60                                                                                                           & 0.769                                                                                     & 0.652                                                                                          \\ \cline{4-6} 
                                         &                                                                                              &                                                                               & 80                                                                                                           & 0.801                                                                                     & 0.684                                                                                          \\ \cline{4-6} 
                                         &                                                                                              &                                                                               & 100                                                                                                          & \textbf{0.814}                                                                            & 0.674                                                                                          \\ \cline{4-6} 
                                         &                                                                                              & \multirow{-4}{*}{49}                                                          & 120                                                                                                          & 0.808                                                                                     & 0.693                                                                                          \\ \cline{3-6} 
                                         &                                                                                              &                                                                               & 60                                                                                                           & 0.776                                                                                     & 0.681                                                                                          \\ \cline{4-6} 
                                         &                                                                                              &                                                                               & 80                                                                                                           & 0.769                                                                                     & \textbf{0.722}                                                                                 \\ \cline{4-6} 
                                         &                                                                                              &                                                                               & 100                                                                                                          & 0.775                                                                                     & 0.658                                                                                          \\ \cline{4-6} 
\multirow{-8}{*}{BreastMNIST}            & \multirow{-8}{*}{4}                                                                          & \multirow{-4}{*}{64}                                                          & 120                                                                                                          & 0.794                                                                                     & 0.694                                                                                          \\ \hline
                                         &                                                                                              &                                                                               & 60                                                                                                           & 0.433                                                                                     & 0.657                                                                                          \\ \cline{4-6} 
                                         &                                                                                              &                                                                               & 80                                                                                                           & 0.41                                                                                      & 0.656                                                                                          \\ \cline{4-6} 
                                         &                                                                                              &                                                                               & 100                                                                                                          & 0.404                                                                                     & 0.648                                                                                          \\ \cline{4-6} 
                                         &                                                                                              & \multirow{-4}{*}{49}                                                          & 120                                                                                                          & 0.415                                                                                     & 0.657                                                                                          \\ \cline{3-6} 
                                         &                                                                                              &                                                                               & 60                                                                                                           & 0.388                                                                                     & 0.628                                                                                          \\ \cline{4-6} 
                                         &                                                                                              &                                                                               & 80                                                                                                           & 0.409                                                                                     & 0.631                                                                                          \\ \cline{4-6} 
                                         &                                                                                              &                                                                               & 100                                                                                                          & 0.423                                                                                     & 0.651                                                                                          \\ \cline{4-6} 
                                         & \multirow{-8}{*}{4}                                                                          & \multirow{-4}{*}{64}                                                          & 120                                                                                                          & 0.413                                                                                     & 0.648                                                                                          \\ \cline{2-6} 
                                         &                                                                                              &                                                                               & 60                                                                                                           & 0.387                                                                                     & 0.62                                                                                           \\ \cline{4-6} 
                                         &                                                                                              &                                                                               & 80                                                                                                           & 0.379                                                                                     & 0.623                                                                                          \\ \cline{4-6} 
                                         &                                                                                              &                                                                               & 100                                                                                                          & \textbf{0.443}                                                                            & \textbf{0.659}                                                                                 \\ \cline{4-6} 
                                         &                                                                                              & \multirow{-4}{*}{49}                                                          & 120                                                                                                          & 0.441                                                                                     & 0.652                                                                                          \\ \cline{3-6} 
                                         &                                                                                              &                                                                               & 60                                                                                                           & 0.393                                                                                     & 0.621                                                                                          \\ \cline{4-6} 
                                         &                                                                                              &                                                                               & 80                                                                                                           & 0.36                                                                                      & 0.593                                                                                          \\ \cline{4-6} 
                                         &                                                                                              &                                                                               & 100                                                                                                          & 0.398                                                                                     & 0.631                                                                                          \\ \cline{4-6} 
\multirow{-16}{*}{OCTMNIST}              & \multirow{-8}{*}{8}                                                                          & \multirow{-4}{*}{64}                                                          & 120                                                                                                          & 0.379                                                                                     & 0.628                                                                                          \\ \hline
                                         &                                                                                              &                                                                               & 60                                                                                                           & 0.4775                                                                                    & 0.6268                                                                                         \\ \cline{4-6} 
                                         &                                                                                              &                                                                               & 80                                                                                                           & 0.48                                                                                      & 0.6646                                                                                         \\ \cline{4-6} 
                                         &                                                                                              &                                                                               & 100                                                                                                          & 0.485                                                                                     & 0.6434                                                                                         \\ \cline{4-6} 
                                         &                                                                                              & \multirow{-4}{*}{49}                                                          & 120                                                                                                          & 0.5                                                                                       & 0.6458                                                                                         \\ \cline{3-6} 
                                         &                                                                                              &                                                                               & 60                                                                                                           & 0.5025                                                                                    & \textbf{0.6648}                                                                                \\ \cline{4-6} 
                                         &                                                                                              &                                                                               & 80                                                                                                           & 0.505                                                                                     & 0.6474                                                                                         \\ \cline{4-6} 
                                         &                                                                                              &                                                                               & 100                                                                                                          & 0.5025                                                                                    & 0.6586                                                                                         \\ \cline{4-6} 
\multirow{-8}{*}{RetinaMNIST}            & \multirow{-8}{*}{5}                                                                          & \multirow{-4}{*}{64}                                                          & 120                                                                                                          & \cellcolor[HTML]{FFFFFF}\textbf{0.5175}                                                   & 0.6559                                                                                         \\ \hline
\end{tabular}
\end{table}

\begin{table}[!ht]
\centering
\caption{Performance analysis of QML framework on simulator based on classification accuracy (ACC) and area under the curve (AUC) metrics under different configurations of quantum models.}
\label{table3}
\begin{tabular}{|c|c|c|c|c|c|}
\hline
\cellcolor[HTML]{FFFFFF}\textbf{Dataset} & \cellcolor[HTML]{FFFFFF}\textbf{\begin{tabular}[c]{@{}c@{}}Number of \\ qubits\end{tabular}} & \textbf{\begin{tabular}[c]{@{}c@{}}Number of\\ embedding gates\end{tabular}} & \cellcolor[HTML]{FFFFFF}\textbf{\begin{tabular}[c]{@{}c@{}}Number of\\ trainable parameters\end{tabular}} & \cellcolor[HTML]{FFFFFF}\textbf{\begin{tabular}[c]{@{}c@{}}Noiseless \\ ACC\end{tabular}} & \cellcolor[HTML]{FFFFFF}\textbf{\begin{tabular}[c]{@{}c@{}}Noiseless\\ AUC\end{tabular}} \\ \hline
                                         &                                                                                              &                                                                              & 60                                                                                                        & 0.6688                                                                                    & 0.6873                                                                                   \\ \cline{4-6} 
                                         &                                                                                              &                                                                              & 80                                                                                                        & 0.6698                                                                                    & 0.7144                                                                                   \\ \cline{4-6} 
                                         &                                                                                              &                                                                              & 100                                                                                                       & 0.6698                                                                                    & 0.7013                                                                                   \\ \cline{4-6} 
                                         &                                                                                              & \multirow{-4}{*}{49}                                                         & 120                                                                                                       & 0.6693                                                                                    & 0.7059                                                                                   \\ \cline{3-6} 
                                         &                                                                                              &                                                                              & 60                                                                                                        & 0.6723                                                                                    & 0.6938                                                                                   \\ \cline{4-6} 
                                         &                                                                                              &                                                                              & 80                                                                                                        & 0.6693                                                                                    & 0.7094                                                                                   \\ \cline{4-6} 
                                         &                                                                                              &                                                                              & 100                                                                                                       & 0.6728                                                                                    & 0.6959                                                                                   \\ \cline{4-6} 
\multirow{-8}{*}{DermaMNIST}             & \multirow{-8}{*}{7}                                                                          & \multirow{-4}{*}{64}                                                         & 120                                                                                                       & \textbf{0.6738}                                                                           & \textbf{0.7343}                                                                          \\ \hline
                                         &                                                                                              &                                                                              & 60                                                                                                        & 0.415                                                                                     & 0.774                                                                                    \\ \cline{4-6} 
                                         &                                                                                              &                                                                              & 80                                                                                                        & 0.371                                                                                     & 0.781                                                                                    \\ \cline{4-6} 
                                         &                                                                                              &                                                                              & 100                                                                                                       & 0.462                                                                                     & 0.811                                                                                    \\ \cline{4-6} 
                                         &                                                                                              & \multirow{-4}{*}{49}                                                         & 120                                                                                                       & 0.432                                                                                     & 0.782                                                                                    \\ \cline{3-6} 
                                         &                                                                                              &                                                                              & 60                                                                                                        & 0.44                                                                                      & 0.792                                                                                    \\ \cline{4-6} 
                                         &                                                                                              &                                                                              & 80                                                                                                        & 0.434                                                                                     & 0.796                                                                                    \\ \cline{4-6} 
                                         &                                                                                              &                                                                              & 100                                                                                                       & 0.421                                                                                     & 0.791                                                                                    \\ \cline{4-6} 
\multirow{-8}{*}{BloodMNIST}             & \multirow{-8}{*}{8}                                                                          & \multirow{-4}{*}{64}                                                         & 120                                                                                                       & \textbf{0.482}                                                                            & \textbf{0.821}                                                                           \\ \hline
                                         &                                                                                              &                                                                              & 60                                                                                                        & 0.3398                                                                                    & 0.7542                                                                                   \\ \cline{4-6} 
                                         &                                                                                              &                                                                              & 80                                                                                                        & 0.3553                                                                                    & 0.7492                                                                                   \\ \cline{4-6} 
                                         &                                                                                              &                                                                              & 100                                                                                                       & 0.3467                                                                                    & 0.7703                                                                                   \\ \cline{4-6} 
                                         &                                                                                              & \multirow{-4}{*}{49}                                                         & 120                                                                                                       & 0.3614                                                                                    & 0.7796                                                                                   \\ \cline{3-6} 
                                         &                                                                                              &                                                                              & 60                                                                                                        & 0.3443                                                                                    & 0.7574                                                                                   \\ \cline{4-6} 
                                         &                                                                                              &                                                                              & 80                                                                                                        & 0.3341                                                                                    & 0.7708                                                                                   \\ \cline{4-6} 
                                         &                                                                                              &                                                                              & 100                                                                                                       & 0.3558                                                                                    & 0.7682                                                                                   \\ \cline{4-6} 
\multirow{-8}{*}{PathMNIST}              & \multirow{-8}{*}{9}                                                                          & \multirow{-4}{*}{64}                                                         & 120                                                                                                       & \textbf{0.3723}                                                                           & \textbf{0.784}                                                                           \\ \hline
                                         &                                                                                              &                                                                              & 60                                                                                                        & 0.3735                                                                                    & 0.7492                                                                                   \\ \cline{4-6} 
                                         &                                                                                              &                                                                              & 80                                                                                                        & 0.3711                                                                                    & 0.7585                                                                                   \\ \cline{4-6} 
                                         &                                                                                              &                                                                              & 100                                                                                                       & 0.369                                                                                     & 0.7624                                                                                   \\ \cline{4-6} 
                                         &                                                                                              & \multirow{-4}{*}{49}                                                         & 120                                                                                                       & 0.3544                                                                                    & 0.7689                                                                                   \\ \cline{3-6} 
                                         &                                                                                              &                                                                              & 60                                                                                                        & 0.3353                                                                                    & 0.729                                                                                    \\ \cline{4-6} 
                                         &                                                                                              &                                                                              & 80                                                                                                        & 0.3528                                                                                    & 0.7709                                                                                   \\ \cline{4-6} 
                                         &                                                                                              &                                                                              & 100                                                                                                       & \textbf{0.3786}                                                                           & 0.7599                                                                                   \\ \cline{4-6} 
\multirow{-8}{*}{OrganSMNIST}            & \multirow{-8}{*}{11}                                                                         & \multirow{-4}{*}{64}                                                         & 120                                                                                                       & 0.3681                                                                                    & \textbf{0.7802}                                                                          \\ \hline
\end{tabular}
\end{table}

\subsection{Noiseless results}

We present the noiseless results of the QML framework using a simulator based on ACC and AUC metrics under different settings of input embedding features and quantum circuit trainable parameters for the MedMNIST datasets we explored in Tables \ref{table2} and \ref{table3}. The QML evaluation is performed by considering all the training and testing samples of the MedMNIST datasets as given in Table \ref{table1}. The bold values in Tables \ref{table2} and \ref{table3} represent the best accuracy and AUC values. For the PneumoniaMNIST dataset, our QML model with 4 qubits, 49 input embedding features, and 120 trainable parameters provides the highest classification accuracy of 85.26\% and AUC of 81.97\% as compared to the other settings as shown in Table \ref{table2}. In the case of the BreastMNIST dataset, we got the best accuracy of 81.40\% for the 4-qubit quantum model having 49 embedding features and 100 trainable parameters, while the highest AUC of 72.20\% is achieved for the model with 64 embedding features and 80 trainable parameters. Similarly, we got the highest accuracy of 44.30\% and AUC of 65.90\% for an 8-qubit quantum model having 49 embedding features and 100 trainable parameters for the OCTMNIST dataset. For the RetinaMNIST dataset, we achieved the best classification accuracy of 51.75\% for the 5-qubit quantum model having 64 input embedding features and 120 trainable parameters. The highest AUC of 66.48\% is obtained for the quantum model having 64 embedding features and 60 trainable parameters as depicted in Table \ref{table2}. A 7-qubit quantum model with 64 input embedding features and 120 trainable parameters provides the highest accuracy of 67.38\% and AUC of 73.43\% for the 7-class DermaMNIST dataset as shown in Table \ref{table3}. For an 8-class BloodMNIST dataset, we achieved the highest accuracy of 48.20\% and AUC of 82.10\% using the QML model with 64 embedding features and 120 trainable parameters. Similarly, we got the best accuracy of 37.23\% and AUC of 78.40\% for a 9-class PathMNIST dataset using the QML model having 64 embedding features and 120 trainable parameters. Lastly, the QML model with 64 features and 100 trainable parameters provides the best classification accuracy of 37.86\% for the 11-class OrganSMNIST dataset, while the best AUC of 78.02\% is achieved for the QML model having 64 embedding features and 120 trainable parameters. It is clear from these results that the quantum models without any classical neural networks and with very limited input features show promising capability to classify medical imaging datasets. Also, it is observed from the Tables \ref{table2} and \ref{table3} that there is a significant difference between the accuracy and AUC values, for instance, we got an accuracy of 48.20\% and AUC of 82.10\% for the BloodMNIST dataset. A similar trend can be seen for the PathMNIST and OrganSMNIST datasets. Note that we considered AUC as our primary performance evaluation metric because it is less sensitive to the class imbalance of the MedMNIST datasets as compared to the accuracy metric. Moreover, it can be noted that for most of the multi-class datasets, we got the best results corresponding to the QML models having more embedding features and trainable parameters in the quantum circuit.   

\subsection{Ablation study of error suppression and mitigation}

We also performed an ablation study on IBM Cleveland to understand the impact of error suppression and error mitigation techniques on the performance of QML models for medical imaging classification as shown in Table \ref{table4}. We performed this ablation on the OCTMNIST dataset because of the significant effect of hardware noise on the performance of the QML model. For the OCTMNIST dataset, our QML workflow provides the noiseless accuracy of 44.30\% and AUC of 65.90\% as shown in Table \ref{table2}. When we ran this quantum model on real IBM hardware without any error suppression and mitigation techniques, we got an ACC of 28.60\% and an AUC of 54.50\%. Therefore, we observed a reduction of 15.70\% and 11.40\% in accuracy and AUC, respectively due to the noise of IBM Cleveland hardware. In this ablation, since we are using a sampler in the inference stage as shown in Fig. \ref{figure1}, we considered two error suppression techniques i.e., dynamical decoupling and gate twirling and one error mitigation technique i.e., M3 mitigation. It is observed from the Table \ref{table4} that when we use both error suppression techniques and M3 mitigation together (DD+Twirling+M3), we got the best accuracy of 39.70\% and AUC of 62.20\%, thereby providing an improvement of 11.10\% and 7.7\% in accuracy and AUC respectively when compared to the case where no suppression and mitigation is used. Based on this ablation study on the OCTMNIST dataset, we used this combination (DD+Twirling+M3) for all the MedMNIST datasets we studied.

\begin{table}[!ht]
\centering
\caption{Ablation of different error suppression/mitigation techniques using OCTMNIST dataset on real IBM Cleveland hardware.}
\label{table4}
\begin{tabular}{|c|c|c|}
\hline
                                & IBM Cleveland (ACC) & IBM Cleveland (AUC) \\ \hline
No error suppression/mitigation & 0.286               & 0.545               \\ \hline
Error suppression (DD+Twirling)                     & 0.33                & 0.591               \\ \hline
M3 error mitigation                   & 0.285               & 0.533               \\ \hline
DD+Twirling+M3                  & \textbf{0.397}      & \textbf{0.622}      \\ \hline
\end{tabular}
\end{table}

\subsection{IBM Cleveland hardware results}

We selected the best performing QML models from Tables \ref{table2} and \ref{table3} based on the AUC metric for inference on IBM Cleveland for all considered MedMNIST datasets. Due to hardware limitations, we considered 1000 testing samples to evaluate the QML models for the datasets having a large number of testing samples. We presented noiseless and IBM hardware results (with and without error suppression and mitigation) together in Table \ref{table5}  along with the specifications of the quantum models such as number of qubits, number of embedded features/gates, trainable parameters, quantum circuit gates, and circuit depth. It is observed from Table \ref{table5} that even with the small number of input embedding features, our QML models consistently provide promising results in terms of classification accuracy and AUC metrics. It is worth noting that the noiseless results and hardware results (without error suppression and mitigation) are quite comparable for most of the MedMNIST datasets except the OCTMNIST and OrganSMNIST datasets. This is because we generated device-aware quantum circuits with optimal circuit gates, depth, and robustness against hardware noise. For the OCTMNIST dataset, the classification accuracy and AUC on the hardware (without error suppression and mitigation) are reduced by 15.70\% and 11.40\% respectively as compared to the noiseless results. Similarly, accuracy and AUC decreased by 14.11\% and 4.8\% respectively for the OrganSMNIST dataset as compared to the noiseless results. This indicates that there is scope for further improvement in the design of device-aware quantum circuits. To mitigate the effect of hardware noise, we ran the hardware experiments for the selected MedMNIST datasets using an error suppression and error mitigation combination (DD+Twirling+M3) based on the ablation study conducted in Table \ref{table4}. We noted improvements in classification accuracy and AUC results using the DD+Twirling+M3 combination as compared to the hardware results without error suppression and mitigation for all the datasets except the BreastMNIST dataset. For instance, the classification accuracy and AUC values are significantly improved by 11.10\% and 7.7\% respectively as compared to the hardware results without error suppression and mitigation for the OCTMNIST dataset. For the BreastMNIST dataset, the accuracy and AUC values decreased by 1.3\% and 2.4\% respectively when compared to the hardware results without error suppression and mitigation.  

\begin{table}[!ht]
\centering
\caption{Performance analysis of QML framework on real IBM Cleveland hardware based on classification accuracy (ACC) and area under
the curve (AUC) metrices for different MedMNIST datasets.}
\label{table5}
\scalebox{0.86}{
\begin{tabular}{|c|c|c|c|c|c|cc|cccc|}
\hline
\multirow{3}{*}{\textbf{Dataset}} & \multirow{3}{*}{\textbf{Qubits}} & \multirow{3}{*}{\textbf{\begin{tabular}[c]{@{}c@{}}Embed\\ gates\end{tabular}}} & \multirow{3}{*}{\textbf{\begin{tabular}[c]{@{}c@{}}Tunable\\ params\end{tabular}}} & \multirow{3}{*}{\textbf{\begin{tabular}[c]{@{}c@{}}Circuit\\ gates\end{tabular}}} & \multirow{3}{*}{\textbf{\begin{tabular}[c]{@{}c@{}}Circuit\\ depth\end{tabular}}} & \multicolumn{2}{c|}{\multirow{2}{*}{\textbf{Noiseless}}} & \multicolumn{4}{c|}{\textbf{IBM Cleveland}}                                                      \\ \cline{9-12} 
                                  &                                  &                                                                                 &                                                                                    &                                                                                   &                                                                                   & \multicolumn{2}{c|}{}                                    & \multicolumn{2}{c|}{No supp./miti.}                       & \multicolumn{2}{c|}{DD+Twir+M3}      \\ \cline{7-12} 
                                  &                                  &                                                                                 &                                                                                    &                                                                                   &                                                                                   & \multicolumn{1}{c|}{ACC}              & AUC              & \multicolumn{1}{c|}{ACC}    & \multicolumn{1}{c|}{AUC}    & \multicolumn{1}{c|}{ACC}    & AUC    \\ \hline
PneumoniaMNIST                         & 4                                & 49                                                                              & 120                                                                                & 184                                                                               & 51                                                                                & \multicolumn{1}{c|}{0.8526}           & 0.8197           & \multicolumn{1}{c|}{0.8381} & \multicolumn{1}{c|}{0.8038} & \multicolumn{1}{c|}{0.8542} & 0.8218 \\ \hline
BreastMNIST                            & 4                                & 64                                                                              & 80                                                                                 & 187                                                                                & 59                                                                                & \multicolumn{1}{c|}{0.769}            & 0.722            & \multicolumn{1}{c|}{0.782}  & \multicolumn{1}{c|}{0.738}  & \multicolumn{1}{c|}{0.769}  & 0.714  \\ \hline
OCTMNIST                               & 8                                & 49                                                                              & 100                                                                                & 208                                                                                & 47                                                                                & \multicolumn{1}{c|}{0.443}            & 0.659            & \multicolumn{1}{c|}{0.286}  & \multicolumn{1}{c|}{0.545}  & \multicolumn{1}{c|}{0.397}  & 0.622  \\ \hline
RetinaMNIST                            & 5                                & 64                                                                              & 60                                                                                 & 140                                                                               & 29                                                                                & \multicolumn{1}{c|}{0.5025}           & 0.6648           & \multicolumn{1}{c|}{0.5025} & \multicolumn{1}{c|}{0.6638} & \multicolumn{1}{c|}{0.505}  & 0.6642 \\ \hline
DermaMNIST                             & 7                                & 64                                                                              & 120                                                                                & 212                                                                               & 33                                                                                & \multicolumn{1}{c|}{0.6738}           & 0.7343           & \multicolumn{1}{c|}{0.671}  & \multicolumn{1}{c|}{0.6825} & \multicolumn{1}{c|}{0.677}  & 0.6982 \\ \hline
BloodMNIST                             & 8                                & 64                                                                              & 120                                                                                & 211                                                                                & 27                                                                                & \multicolumn{1}{c|}{0.482}            & 0.821            & \multicolumn{1}{c|}{0.422}  & \multicolumn{1}{c|}{0.824}  & \multicolumn{1}{c|}{0.424}  & 0.826  \\ \hline
PathMNIST                              & 9                                & 64                                                                              & 120                                                                                & 218                                                                               & 27                                                                                & \multicolumn{1}{c|}{0.3723}           & 0.784            & \multicolumn{1}{c|}{0.35}   & \multicolumn{1}{c|}{0.7544} & \multicolumn{1}{c|}{0.418}  & 0.7795 \\ \hline
OrganSMNIST                            & 11                               & 64                                                                              & 120                                                                                & 213                                                                               & 23                                                                                & \multicolumn{1}{c|}{0.3681}           & 0.7802           & \multicolumn{1}{c|}{0.227}  & \multicolumn{1}{c|}{0.7322} & \multicolumn{1}{c|}{0.34}   & 0.7678 \\ \hline
\end{tabular}}
\end{table}

\begin{table}[!hb]
\centering
\caption{Comparative analysis of QML model and classical benchmark ML models using AUC and ACC metrices on MedMNIST datasets.}
\label{table6}
\begin{tabular}{ccccccccc}
\hline
\rowcolor[HTML]{FFFFFF} 
\multicolumn{1}{|c|}{\cellcolor[HTML]{FFFFFF}}                                                               & \multicolumn{2}{c|}{\cellcolor[HTML]{FFFFFF}\textbf{PneumoniaMNIST}}                                                      & \multicolumn{2}{c|}{\cellcolor[HTML]{FFFFFF}\textbf{BreastMNIST}}                                                         & \multicolumn{2}{c|}{\cellcolor[HTML]{FFFFFF}\textbf{OCTMNIST}}                                                            & \multicolumn{2}{c|}{\cellcolor[HTML]{FFFFFF}\textbf{RetinaMNIST}}                                                         \\ \cline{2-9} 
\rowcolor[HTML]{FFFFFF} 
\multicolumn{1}{|c|}{\multirow{-2}{*}{\cellcolor[HTML]{FFFFFF}\textbf{Method (Input size)}}}                              & \multicolumn{1}{c|}{\cellcolor[HTML]{FFFFFF}AUC}            & \multicolumn{1}{c|}{\cellcolor[HTML]{FFFFFF}ACC}            & \multicolumn{1}{c|}{\cellcolor[HTML]{FFFFFF}AUC}            & \multicolumn{1}{c|}{\cellcolor[HTML]{FFFFFF}ACC}            & \multicolumn{1}{c|}{\cellcolor[HTML]{FFFFFF}AUC}            & \multicolumn{1}{c|}{\cellcolor[HTML]{FFFFFF}ACC}            & \multicolumn{1}{c|}{\cellcolor[HTML]{FFFFFF}AUC}            & \multicolumn{1}{c|}{\cellcolor[HTML]{FFFFFF}ACC}            \\ \hline
\rowcolor[HTML]{FFFFFF} 
\multicolumn{1}{|c|}{\cellcolor[HTML]{FFFFFF}ResNet-18 (28$\times$28$\times$3) \cite{he2016deep}}                                                 & \multicolumn{1}{c|}{\cellcolor[HTML]{FFFFFF}0.944}          & \multicolumn{1}{c|}{\cellcolor[HTML]{FFFFFF}0.854}          & \multicolumn{1}{c|}{\cellcolor[HTML]{FFFFFF}0.901}          & \multicolumn{1}{c|}{\cellcolor[HTML]{FFFFFF}\textbf{0.863}} & \multicolumn{1}{c|}{\cellcolor[HTML]{FFFFFF}0.943}          & \multicolumn{1}{c|}{\cellcolor[HTML]{FFFFFF}0.743}          & \multicolumn{1}{c|}{\cellcolor[HTML]{FFFFFF}0.717}          & \multicolumn{1}{c|}{\cellcolor[HTML]{FFFFFF}0.524}          \\ \hline
\rowcolor[HTML]{FFFFFF} 
\multicolumn{1}{|c|}{\cellcolor[HTML]{FFFFFF}ResNet-18 (224$\times$224$\times$3) \cite{he2016deep}}                                                & \multicolumn{1}{c|}{\cellcolor[HTML]{FFFFFF}0.956}          & \multicolumn{1}{c|}{\cellcolor[HTML]{FFFFFF}0.864}          & \multicolumn{1}{c|}{\cellcolor[HTML]{FFFFFF}0.891}          & \multicolumn{1}{c|}{\cellcolor[HTML]{FFFFFF}0.833}          & \multicolumn{1}{c|}{\cellcolor[HTML]{FFFFFF}0.958}          & \multicolumn{1}{c|}{\cellcolor[HTML]{FFFFFF}0.763}          & \multicolumn{1}{c|}{\cellcolor[HTML]{FFFFFF}0.71}           & \multicolumn{1}{c|}{\cellcolor[HTML]{FFFFFF}0.493}          \\ \hline
\rowcolor[HTML]{FFFFFF} 
\multicolumn{1}{|c|}{\cellcolor[HTML]{FFFFFF}ResNet-50 (28$\times$28$\times$3) \cite{he2016deep}}                                                 & \multicolumn{1}{c|}{\cellcolor[HTML]{FFFFFF}0.948}          & \multicolumn{1}{c|}{\cellcolor[HTML]{FFFFFF}0.854}          & \multicolumn{1}{c|}{\cellcolor[HTML]{FFFFFF}0.857}          & \multicolumn{1}{c|}{\cellcolor[HTML]{FFFFFF}0.812}          & \multicolumn{1}{c|}{\cellcolor[HTML]{FFFFFF}0.952}          & \multicolumn{1}{c|}{\cellcolor[HTML]{FFFFFF}0.762}          & \multicolumn{1}{c|}{\cellcolor[HTML]{FFFFFF}0.726}          & \multicolumn{1}{c|}{\cellcolor[HTML]{FFFFFF}0.528}          \\ \hline
\rowcolor[HTML]{FFFFFF} 
\multicolumn{1}{|c|}{\cellcolor[HTML]{FFFFFF}ResNet-50 (224$\times$224$\times$3) \cite{he2016deep}}                                                & \multicolumn{1}{c|}{\cellcolor[HTML]{FFFFFF}0.962}          & \multicolumn{1}{c|}{\cellcolor[HTML]{FFFFFF}0.884}          & \multicolumn{1}{c|}{\cellcolor[HTML]{FFFFFF}0.866}          & \multicolumn{1}{c|}{\cellcolor[HTML]{FFFFFF}0.842}          & \multicolumn{1}{c|}{\cellcolor[HTML]{FFFFFF}0.958}          & \multicolumn{1}{c|}{\cellcolor[HTML]{FFFFFF}\textbf{0.776}} & \multicolumn{1}{c|}{\cellcolor[HTML]{FFFFFF}0.716}          & \multicolumn{1}{c|}{\cellcolor[HTML]{FFFFFF}0.511}          \\ \hline
\rowcolor[HTML]{FFFFFF} 
\multicolumn{1}{|c|}{\cellcolor[HTML]{FFFFFF}Auto-sklearn (28$\times$28$\times$3) \cite{Feurer2019}}                                                   & \multicolumn{1}{c|}{\cellcolor[HTML]{FFFFFF}0.942}          & \multicolumn{1}{c|}{\cellcolor[HTML]{FFFFFF}0.855}          & \multicolumn{1}{c|}{\cellcolor[HTML]{FFFFFF}0.836}          & \multicolumn{1}{c|}{\cellcolor[HTML]{FFFFFF}0.803}          & \multicolumn{1}{c|}{\cellcolor[HTML]{FFFFFF}0.887}          & \multicolumn{1}{c|}{\cellcolor[HTML]{FFFFFF}0.601}          & \multicolumn{1}{c|}{\cellcolor[HTML]{FFFFFF}0.69}           & \multicolumn{1}{c|}{\cellcolor[HTML]{FFFFFF}0.515}          \\ \hline
\rowcolor[HTML]{FFFFFF} 
\multicolumn{1}{|c|}{\cellcolor[HTML]{FFFFFF}AutoKeras (28$\times$28$\times$3) \cite{jin2019auto}}                                                      & \multicolumn{1}{c|}{\cellcolor[HTML]{FFFFFF}0.947}          & \multicolumn{1}{c|}{\cellcolor[HTML]{FFFFFF}0.878}          & \multicolumn{1}{c|}{\cellcolor[HTML]{FFFFFF}0.871}          & \multicolumn{1}{c|}{\cellcolor[HTML]{FFFFFF}0.831}          & \multicolumn{1}{c|}{\cellcolor[HTML]{FFFFFF}0.955}          & \multicolumn{1}{c|}{\cellcolor[HTML]{FFFFFF}0.763}          & \multicolumn{1}{c|}{\cellcolor[HTML]{FFFFFF}0.719}          & \multicolumn{1}{c|}{\cellcolor[HTML]{FFFFFF}0.503}          \\ \hline
\rowcolor[HTML]{FFFFFF} 
\multicolumn{1}{|c|}{\cellcolor[HTML]{FFFFFF}\begin{tabular}[c]{@{}c@{}}Google AutoML\\ Vision (28$\times$28$\times$3) \cite{google_automl_vision}\end{tabular}} & \multicolumn{1}{c|}{\cellcolor[HTML]{FFFFFF}\textbf{0.991}} & \multicolumn{1}{c|}{\cellcolor[HTML]{FFFFFF}\textbf{0.946}} & \multicolumn{1}{c|}{\cellcolor[HTML]{FFFFFF}\textbf{0.919}} & \multicolumn{1}{c|}{\cellcolor[HTML]{FFFFFF}0.861}          & \multicolumn{1}{c|}{\cellcolor[HTML]{FFFFFF}\textbf{0.963}} & \multicolumn{1}{c|}{\cellcolor[HTML]{FFFFFF}0.771}          & \multicolumn{1}{c|}{\cellcolor[HTML]{FFFFFF}\textbf{0.75}}  & \multicolumn{1}{c|}{\cellcolor[HTML]{FFFFFF}\textbf{0.531}} \\ \hline
\multicolumn{1}{|c|}{\begin{tabular}[c]{@{}c@{}}\textbf{QML model} (7$\times$7 or 8$\times$8) \\ (IBM Cleveland)\end{tabular}}           & \multicolumn{1}{c|}{0.822}                                  & \multicolumn{1}{c|}{0.854}                                  & \multicolumn{1}{c|}{0.714}                                  & \multicolumn{1}{c|}{0.769}                                  & \multicolumn{1}{c|}{0.622}                                  & \multicolumn{1}{c|}{0.397}                                  & \multicolumn{1}{c|}{0.664}                                  & \multicolumn{1}{c|}{0.505}                                  \\ \hline
                                                                                                             &                                                             &                                                             &                                                             &                                                             &                                                             &                                                             &                                                             &                                                             \\ \hline
\rowcolor[HTML]{FFFFFF} 
\multicolumn{1}{|c|}{\cellcolor[HTML]{FFFFFF}}                                                               & \multicolumn{2}{c|}{\cellcolor[HTML]{FFFFFF}\textbf{DermaMNIST}}                                                          & \multicolumn{2}{c|}{\cellcolor[HTML]{FFFFFF}\textbf{BloodMNIST}}                                                          & \multicolumn{2}{c|}{\cellcolor[HTML]{FFFFFF}\textbf{PathMNIST}}                                                           & \multicolumn{2}{c|}{\cellcolor[HTML]{FFFFFF}\textbf{OrganSMNIST}}                                                         \\ \cline{2-9} 
\rowcolor[HTML]{FFFFFF} 
\multicolumn{1}{|c|}{\multirow{-2}{*}{\cellcolor[HTML]{FFFFFF}\textbf{Method (Input size)}}}                              & \multicolumn{1}{c|}{\cellcolor[HTML]{FFFFFF}AUC}            & \multicolumn{1}{c|}{\cellcolor[HTML]{FFFFFF}ACC}            & \multicolumn{1}{c|}{\cellcolor[HTML]{FFFFFF}AUC}            & \multicolumn{1}{c|}{\cellcolor[HTML]{FFFFFF}ACC}            & \multicolumn{1}{c|}{\cellcolor[HTML]{FFFFFF}AUC}            & \multicolumn{1}{c|}{\cellcolor[HTML]{FFFFFF}ACC}            & \multicolumn{1}{c|}{\cellcolor[HTML]{FFFFFF}AUC}            & \multicolumn{1}{c|}{\cellcolor[HTML]{FFFFFF}ACC}            \\ \hline
\rowcolor[HTML]{FFFFFF} 
\multicolumn{1}{|c|}{\cellcolor[HTML]{FFFFFF}ResNet-18 (28$\times$28$\times$3) \cite{he2016deep}}                                                 & \multicolumn{1}{c|}{\cellcolor[HTML]{FFFFFF}0.917}          & \multicolumn{1}{c|}{\cellcolor[HTML]{FFFFFF}0.735}          & \multicolumn{1}{c|}{\cellcolor[HTML]{FFFFFF}\textbf{0.998}} & \multicolumn{1}{c|}{\cellcolor[HTML]{FFFFFF}0.958}          & \multicolumn{1}{c|}{\cellcolor[HTML]{FFFFFF}0.983}          & \multicolumn{1}{c|}{\cellcolor[HTML]{FFFFFF}0.907}          & \multicolumn{1}{c|}{\cellcolor[HTML]{FFFFFF}0.972}          & \multicolumn{1}{c|}{\cellcolor[HTML]{FFFFFF}0.782}          \\ \hline
\rowcolor[HTML]{FFFFFF} 
\multicolumn{1}{|c|}{\cellcolor[HTML]{FFFFFF}ResNet-18 (224$\times$224$\times$3) \cite{he2016deep}}                                                & \multicolumn{1}{c|}{\cellcolor[HTML]{FFFFFF}\textbf{0.92}}  & \multicolumn{1}{c|}{\cellcolor[HTML]{FFFFFF}0.754}          & \multicolumn{1}{c|}{\cellcolor[HTML]{FFFFFF}\textbf{0.998}} & \multicolumn{1}{c|}{\cellcolor[HTML]{FFFFFF}0.963}          & \multicolumn{1}{c|}{\cellcolor[HTML]{FFFFFF}0.989}          & \multicolumn{1}{c|}{\cellcolor[HTML]{FFFFFF}0.909}          & \multicolumn{1}{c|}{\cellcolor[HTML]{FFFFFF}0.974}          & \multicolumn{1}{c|}{\cellcolor[HTML]{FFFFFF}0.778}          \\ \hline
\rowcolor[HTML]{FFFFFF} 
\multicolumn{1}{|c|}{\cellcolor[HTML]{FFFFFF}ResNet-50 (28$\times$28$\times$3) \cite{he2016deep}}                                                 & \multicolumn{1}{c|}{\cellcolor[HTML]{FFFFFF}0.913}          & \multicolumn{1}{c|}{\cellcolor[HTML]{FFFFFF}0.735}          & \multicolumn{1}{c|}{\cellcolor[HTML]{FFFFFF}0.997}          & \multicolumn{1}{c|}{\cellcolor[HTML]{FFFFFF}0.956}          & \multicolumn{1}{c|}{\cellcolor[HTML]{FFFFFF}\textbf{0.99}}  & \multicolumn{1}{c|}{\cellcolor[HTML]{FFFFFF}\textbf{0.911}} & \multicolumn{1}{c|}{\cellcolor[HTML]{FFFFFF}0.972}          & \multicolumn{1}{c|}{\cellcolor[HTML]{FFFFFF}0.77}           \\ \hline
\rowcolor[HTML]{FFFFFF} 
\multicolumn{1}{|c|}{\cellcolor[HTML]{FFFFFF}ResNet-50 (224$\times$224$\times$3) \cite{he2016deep}}                                                & \multicolumn{1}{c|}{\cellcolor[HTML]{FFFFFF}0.912}          & \multicolumn{1}{c|}{\cellcolor[HTML]{FFFFFF}0.731}          & \multicolumn{1}{c|}{\cellcolor[HTML]{FFFFFF}0.997}          & \multicolumn{1}{c|}{\cellcolor[HTML]{FFFFFF}0.95}           & \multicolumn{1}{c|}{\cellcolor[HTML]{FFFFFF}0.989}          & \multicolumn{1}{c|}{\cellcolor[HTML]{FFFFFF}0.892}          & \multicolumn{1}{c|}{\cellcolor[HTML]{FFFFFF}\textbf{0.975}} & \multicolumn{1}{c|}{\cellcolor[HTML]{FFFFFF}0.785}          \\ \hline
\rowcolor[HTML]{FFFFFF} 
\multicolumn{1}{|c|}{\cellcolor[HTML]{FFFFFF}Auto-sklearn (28$\times$28$\times$3) \cite{Feurer2019}}                                                   & \multicolumn{1}{c|}{\cellcolor[HTML]{FFFFFF}0.902}          & \multicolumn{1}{c|}{\cellcolor[HTML]{FFFFFF}0.719}          & \multicolumn{1}{c|}{\cellcolor[HTML]{FFFFFF}0.984}          & \multicolumn{1}{c|}{\cellcolor[HTML]{FFFFFF}0.878}          & \multicolumn{1}{c|}{\cellcolor[HTML]{FFFFFF}0.934}          & \multicolumn{1}{c|}{\cellcolor[HTML]{FFFFFF}0.716}          & \multicolumn{1}{c|}{\cellcolor[HTML]{FFFFFF}0.945}          & \multicolumn{1}{c|}{\cellcolor[HTML]{FFFFFF}0.672}          \\ \hline
\rowcolor[HTML]{FFFFFF} 
\multicolumn{1}{|c|}{\cellcolor[HTML]{FFFFFF}AutoKeras (28$\times$28$\times$3) \cite{jin2019auto}}                                                      & \multicolumn{1}{c|}{\cellcolor[HTML]{FFFFFF}0.915}          & \multicolumn{1}{c|}{\cellcolor[HTML]{FFFFFF}0.749}          & \multicolumn{1}{c|}{\cellcolor[HTML]{FFFFFF}\textbf{0.998}} & \multicolumn{1}{c|}{\cellcolor[HTML]{FFFFFF}0.961}          & \multicolumn{1}{c|}{\cellcolor[HTML]{FFFFFF}0.959}          & \multicolumn{1}{c|}{\cellcolor[HTML]{FFFFFF}0.834}          & \multicolumn{1}{c|}{\cellcolor[HTML]{FFFFFF}0.974}          & \multicolumn{1}{c|}{\cellcolor[HTML]{FFFFFF}\textbf{0.813}} \\ \hline
\rowcolor[HTML]{FFFFFF} 
\multicolumn{1}{|c|}{\cellcolor[HTML]{FFFFFF}\begin{tabular}[c]{@{}c@{}}Google AutoML\\ Vision (28$\times$28$\times$3) \cite{google_automl_vision}\end{tabular}} & \multicolumn{1}{c|}{\cellcolor[HTML]{FFFFFF}0.914}          & \multicolumn{1}{c|}{\cellcolor[HTML]{FFFFFF}\textbf{0.768}} & \multicolumn{1}{c|}{\cellcolor[HTML]{FFFFFF}\textbf{0.998}} & \multicolumn{1}{c|}{\cellcolor[HTML]{FFFFFF}\textbf{0.966}} & \multicolumn{1}{c|}{\cellcolor[HTML]{FFFFFF}0.944}          & \multicolumn{1}{c|}{\cellcolor[HTML]{FFFFFF}0.728}          & \multicolumn{1}{c|}{\cellcolor[HTML]{FFFFFF}0.964}          & \multicolumn{1}{c|}{\cellcolor[HTML]{FFFFFF}0.749}          \\ \hline
\multicolumn{1}{|c|}{\begin{tabular}[c]{@{}c@{}}\textbf{QML model} (7$\times$7 or 8$\times$8) \\ (IBM Cleveland)\end{tabular}}           & \multicolumn{1}{c|}{0.698}                                  & \multicolumn{1}{c|}{0.677}                                  & \multicolumn{1}{c|}{0.826}                                  & \multicolumn{1}{c|}{0.424}                                  & \multicolumn{1}{c|}{0.78}                                   & \multicolumn{1}{c|}{0.418}                                  & \multicolumn{1}{c|}{0.768}                                  & \multicolumn{1}{c|}{0.34}                                   \\ \hline
\end{tabular}
\end{table}

\subsection{Comparison with classical ML models}

We also provided a comparative analysis with the benchmarks of the classical ML models provided in MedMNIST \cite{Yang_2023} as shown in Table \ref{table6}. To the best of our knowledge, this benchmarking study is the first to provide a comprehensive evaluation of QML (without using any classical neural networks) on real IBM quantum hardware using large-scale MedMNIST datasets. Although the classification results obtained by our QML workflow do not outperform the best-performing classical ML model due to the inherent limitations of current NISQ hardware, our results represent a significant step forward in evaluating QML for realistic applications in medical imaging. All classical ML models leverage the full spatial dimensions of input images, whereas our QML model operates on a significantly reduced feature set, dictated by quantum hardware constraints. Despite this, our QML model achieves promising results on IBM Cleveland in terms of classification accuracy and AUC values. For the 2-class PneumoniaMNIST dataset, we obtained good AUC and classification accuracies of 82.2\% and 85.4\% respectively. Our QML accuracy of 85.4\% for the PneumoniaMNIST dataset is equal to the accuracy obtained by the classical ML models including ResNet-18 (28) \cite{he2016deep} and ResNet-50 (28) \cite{he2016deep} as shown in Table \ref{table6}. Also, our QML accuracy 85.4\% for PneumoniaMNIST dataset is lagging by only 0.1\%, 1\%, 2.4\%, and 3\% when compared with the Auto-sklearn \cite{Feurer2019}, ResNet-18 (224) \cite{he2016deep}, AutoKeras \cite{jin2019auto}, and ResNet-50 (224) \cite{he2016deep} classical ML models, respectively. Similarly, our QML model classification accuracy is lagging by 3.4\% when compared to the Auto-sklearn \cite{Feurer2019} classical ML model for the BreastMNIST dataset. Moreover, in the case of the 5-class RetinaMNIST dataset, we achieved an AUC and accuracy of 66.4\% and 50.5\% respectively which lags by 8.6\% and 2.6\% in AUC and accuracy when compared to the best performing classical Google AutoML Vision \cite{google_automl_vision} model. The AUC value 66.4\% of our QML model for RetinaMNIST lags only by 2.6\% compared to the Auto-sklearn \cite{Feurer2019} classical ML model. Importantly, our QML model accuracy of 50.5\% for RetinaMNIST outperforms the ResNet-18 (224) \cite{he2016deep} and AutoKeras \cite{jin2019auto} classical ML models by 1.2\% and 0.2\%, respectively. Even for the 7-class DermaMNIST dataset, our QML model obtained good results with AUC and accuracy values of 69.8\% and 67.7\% respectively. The accuracy value achieved by our QML model for the DermaMNIST dataset only lags by 4.2\% and 9.1\% when compared to the Auto-sklearn \cite{Feurer2019} model and the best performing classical Google AutoML Vision \cite{google_automl_vision} ML model, respectively. Our QML model can achieve an AUC of more than 70\% for all the MedMNIST datasets we explored except for the OCTMNIST, RetinaMNIST, and DermaMNIST datasets. It is worth mentioning that for the 2-class PneumoniaMNIST and 8-class BloodMNIST datasets, we obtained AUCs of more than 80\%. These results underscore the promise of QML in medical image classification, particularly when benchmarked against the existing literature on image classification using quantum approaches, where achieving comparable performance with classical models remains an open challenge. Our study provides a critical evaluation of QML for realistic applications, showcasing its capability even in the current phase of quantum computing. By systematically benchmarking QML performance on IBM Cleveland across diverse MedMNIST datasets, this work paves the way for future advancements in QML and highlights its potential to complement classical ML in the years to come.

To further assess the effectiveness of our QML model, we conducted a comparative analysis by training classical ML baselines, ResNet-18 and ResNet-50, using the same 8×8 input image features as used in our QML model. This ensures a fair and direct comparison of the 5-class RetinaMNIST dataset, as summarized in Table \ref{table7}. We selected the RetinaMNIST dataset for this experiment because it presents a high level of complexity, resulting in the lowest performance among all classical ML models. Despite having significantly fewer trainable parameters (60) compared to ResNet-18 (11.2 million) and ResNet-50 (23.5 million), our QML model demonstrates superior classification performance. Our QML model surpasses ResNet-18 and ResNet-50 by improving classification accuracy by 8\% and 7\%, respectively, while also achieving AUC gains of 5.41\% and 1.54\%, respectively. These results highlight that our QML model achieves quantum advantage in both performance and computational complexity. Motivated by these findings, we aim to further explore advanced embedding techniques to incorporate more features into quantum models, as well as to further optimize ansatz design to enhance expressivity. These improvements will be crucial in demonstrating quantum advantage on a larger scale in future experiments.

\begin{table}[h]
\centering
\caption{Comparative analysis of QML model and classical benchmark ML models by considering same number of input features on 5-class RetinaMNIST dataset.}
\label{table7}
\begin{tabular}{|c|c|cc|}
\hline
\rowcolor[HTML]{FFFFFF} 
\cellcolor[HTML]{FFFFFF}{\color[HTML]{000000} }                                                           & \cellcolor[HTML]{FFFFFF}{\color[HTML]{000000} }                                                & \multicolumn{2}{c|}{\cellcolor[HTML]{FFFFFF}{\color[HTML]{000000} \textbf{RetinaMNIST (5-class)}}}                                \\ \cline{3-4} 
\rowcolor[HTML]{FFFFFF} 
\multirow{-2}{*}{\cellcolor[HTML]{FFFFFF}{\color[HTML]{000000} \textbf{Method (Input size)}}}             & \multirow{-2}{*}{\cellcolor[HTML]{FFFFFF}{\color[HTML]{000000} \textbf{Trainable parameters}}} & \multicolumn{1}{c|}{\cellcolor[HTML]{FFFFFF}{\color[HTML]{000000} ACC}}   & {\color[HTML]{000000} AUC}                            \\ \hline
\cellcolor[HTML]{FFFFFF}{\color[HTML]{000000} ResNet-18 (8$\times$8) \cite{he2016deep}}                                            & {\color[HTML]{000000} 11.2 million}                                                            & \multicolumn{1}{c|}{{\color[HTML]{000000} 0.425}}                         & {\color[HTML]{000000} 0.6101}                         \\ \hline
\cellcolor[HTML]{FFFFFF}{\color[HTML]{000000} ResNet-50 (8$\times$8) \cite{he2016deep}}                                            & {\color[HTML]{000000} 23.5 million}                                                            & \multicolumn{1}{c|}{\cellcolor[HTML]{FFFFFF}{\color[HTML]{000000} 0.435}} & \cellcolor[HTML]{FFFFFF}{\color[HTML]{000000} 0.6488} \\ \hline
{\color[HTML]{000000}\begin{tabular}[c]{@{}c@{}}\textbf{QML model} (8$\times$8)\\ (IBM Cleveland)\end{tabular}} & {\color[HTML]{000000} \textbf{60}}                                                             & \multicolumn{1}{c|}{{\color[HTML]{000000} \textbf{0.505}}}                & {\color[HTML]{000000} \textbf{0.6642}}                \\ \hline
\end{tabular}
\end{table}

\section{Conclusion}

This work demonstrates the potential of QML in the context of medical imaging by benchmarking the MedMNIST datasets on real quantum hardware. We effectively trained and validated noise-resilient quantum circuits using advanced error suppression and mitigation techniques to tackle hardware noise. The experimental results highlight the feasibility of applying QML to complex real-world datasets such as MedMNIST. In the current NISQ era, we can achieve promising results in terms of classification accuracy and AUC values for binary as well as multi-class datasets with a very limited number of input embedding features. For the 2-class PneumoniaMNIST data set, our best-performing QML model provides accuracy and AUC of 85.4\% and 82.2\%, respectively. Similarly, we obtained comparable results with the classical benchmark model in the case of the 5-class RetinaMNIST data set with accuracy and AUC of 50.5\% and 66.4\%, respectively. Even for the 7-class DermaMNIST dataset, our QML model achieves an accuracy of 67.7\% and AUC of 69.8\% showing the capability of QML for medical imaging classification. Importantly, we achieve quantum advantage both in classification performance and computational complexity compared to classical ML baselines when evaluated on the RetinaMNIST dataset using the same number of input features. This study serves as an important benchmark, demonstrating the readiness of quantum computing for practical applications in healthcare. It also underscores the need for further research to enhance scalability, mitigate hardware limitations, and explore novel quantum algorithms. As quantum technology continues to evolve, its integration into medical imaging workflows has the potential to transform healthcare analytics, paving the way for more advanced and efficient diagnostic tools.

\section*{Data availability}

The MedMNIST datasets used in this benchmarking study can be accessed using \url{https://medmnist.com/}

\section*{Code availability}

All the related codes are available at \url{https://github.com/gurinder-hub/QML_MedMNIST}

\section*{Acknowledgments}
The authors gratefully acknowledge financial support from the NIH (GM130641). We also extend our sincere gratitude to Danil Kaliakin (Research Associate, Cleveland Clinic) and Abdullah Ash Saki (Researcher at IBM Quantum) for their valuable insights and support throughout this project. We are also grateful to the authors of Élivágar and Torchquantum for their great work and open-sourcing the code.  

\section*{Author contributions}
G. Singh and H. Jin proposed and design the study of benchmarking MedMNIST datasets using quantum machine learning on real IBM Cleveland hardware. G. Singh and H. Jin performed the experiments under the supervision of K.M. Merz. G. Singh wrote the
manuscript, with inputs and contributions from all authors.

\section*{Competing interests}
The authors declare no competing interests.

\bibliographystyle{unsrt}  

\begin{thebibliography}{10}

\bibitem{Yang_2023}
Jiancheng Yang, Rui Shi, Donglai Wei, Zequan Liu, Lin Zhao, Bilian Ke, Hanspeter Pfister, and Bingbing Ni.
\newblock Medmnist v2 - a large-scale lightweight benchmark for 2d and 3d biomedical image classification.
\newblock {\em Scientific Data}, 10(1), January 2023.

\bibitem{kumar2024medical}
Rakesh Kumar, Pooja Kumbharkar, Sandeep Vanam, and Sanjeev Sharma.
\newblock Medical images classification using deep learning: a survey.
\newblock {\em Multimedia Tools and Applications}, 83(7):19683--19728, 2024.

\bibitem{yue2024medmambavisionmambamedical}
Yubiao Yue and Zhenzhang Li.
\newblock Medmamba: Vision mamba for medical image classification, 2024.

\bibitem{Ma_2024}
Jun Ma, Yuting He, Feifei Li, Lin Han, Chenyu You, and Bo~Wang.
\newblock Segment anything in medical images.
\newblock {\em Nature Communications}, 15(1), January 2024.

\bibitem{zhang2024segmentmodelmedicalimage}
Yichi Zhang, Zhenrong Shen, and Rushi Jiao.
\newblock Segment anything model for medical image segmentation: Current applications and future directions, 2024.

\bibitem{webber2024diffusion}
George Webber and Andrew~J Reader.
\newblock Diffusion models for medical image reconstruction.
\newblock {\em BJR| Artificial Intelligence}, 1(1):ubae013, 2024.

\bibitem{khader2023medicaldiffusiondenoisingdiffusion}
Firas Khader, Gustav Mueller-Franzes, Soroosh~Tayebi Arasteh, Tianyu Han, Christoph Haarburger, Maximilian Schulze-Hagen, Philipp Schad, Sandy Engelhardt, Bettina Baessler, Sebastian Foersch, Johannes Stegmaier, Christiane Kuhl, Sven Nebelung, Jakob~Nikolas Kather, and Daniel Truhn.
\newblock Medical diffusion: Denoising diffusion probabilistic models for 3d medical image generation, 2023.

\bibitem{Liu_2021}
Yunchao Liu, Srinivasan Arunachalam, and Kristan Temme.
\newblock A rigorous and robust quantum speed-up in supervised machine learning.
\newblock {\em Nature Physics}, 17(9):1013–1017, July 2021.

\bibitem{Abbas_2021}
Amira Abbas, David Sutter, Christa Zoufal, Aurelien Lucchi, Alessio Figalli, and Stefan Woerner.
\newblock The power of quantum neural networks.
\newblock {\em Nature Computational Science}, 1(6):403–409, June 2021.

\bibitem{jin2024integratingmachinelearningquantum}
Hongni Jin and Kenneth M.~Merz Jr.
\newblock Integrating machine learning and quantum circuits for proton affinity predictions, 2024.

\bibitem{yamasaki2023advantagequantummachinelearning}
Hayata Yamasaki, Natsuto Isogai, and Mio Murao.
\newblock Advantage of quantum machine learning from general computational advantages, 2023.

\bibitem{xiao2023practical}
Tailong Xiao, Xinliang Zhai, Xiaoyan Wu, Jianping Fan, and Guihua Zeng.
\newblock Practical advantage of quantum machine learning in ghost imaging.
\newblock {\em Communications Physics}, 6(1):171, 2023.

\bibitem{acharya2024quantumerrorcorrectionsurface}
Rajeev Acharya, Laleh Aghababaie-Beni, Igor Aleiner, Trond~I. Andersen, Markus Ansmann, Frank Arute, Kunal Arya, Abraham Asfaw, Nikita Astrakhantsev, Juan Atalaya, Ryan Babbush, Dave Bacon, Brian Ballard, Joseph~C. Bardin, Johannes Bausch, Andreas Bengtsson, Alexander Bilmes, Sam Blackwell, Sergio Boixo, Gina Bortoli, Alexandre Bourassa, Jenna Bovaird, Leon Brill, Michael Broughton, David~A. Browne, Brett Buchea, Bob~B. Buckley, David~A. Buell, Tim Burger, Brian Burkett, Nicholas Bushnell, Anthony Cabrera, Juan Campero, Hung-Shen Chang, Yu~Chen, Zijun Chen, Ben Chiaro, Desmond Chik, Charina Chou, Jahan Claes, Agnetta~Y. Cleland, Josh Cogan, Roberto Collins, Paul Conner, William Courtney, Alexander~L. Crook, Ben Curtin, Sayan Das, Alex Davies, Laura~De Lorenzo, Dripto~M. Debroy, Sean Demura, Michel Devoret, Agustin~Di Paolo, Paul Donohoe, Ilya Drozdov, Andrew Dunsworth, Clint Earle, Thomas Edlich, Alec Eickbusch, Aviv~Moshe Elbag, Mahmoud Elzouka, Catherine Erickson, Lara Faoro, Edward Farhi, Vinicius~S.
  Ferreira, Leslie~Flores Burgos, Ebrahim Forati, Austin~G. Fowler, Brooks Foxen, Suhas Ganjam, Gonzalo Garcia, Robert Gasca, Élie Genois, William Giang, Craig Gidney, Dar Gilboa, Raja Gosula, Alejandro~Grajales Dau, Dietrich Graumann, Alex Greene, Jonathan~A. Gross, Steve Habegger, John Hall, Michael~C. Hamilton, Monica Hansen, Matthew~P. Harrigan, Sean~D. Harrington, Francisco J.~H. Heras, Stephen Heslin, Paula Heu, Oscar Higgott, Gordon Hill, Jeremy Hilton, George Holland, Sabrina Hong, Hsin-Yuan Huang, Ashley Huff, William~J. Huggins, Lev~B. Ioffe, Sergei~V. Isakov, Justin Iveland, Evan Jeffrey, Zhang Jiang, Cody Jones, Stephen Jordan, Chaitali Joshi, Pavol Juhas, Dvir Kafri, Hui Kang, Amir~H. Karamlou, Kostyantyn Kechedzhi, Julian Kelly, Trupti Khaire, Tanuj Khattar, Mostafa Khezri, Seon Kim, Paul~V. Klimov, Andrey~R. Klots, Bryce Kobrin, Pushmeet Kohli, Alexander~N. Korotkov, Fedor Kostritsa, Robin Kothari, Borislav Kozlovskii, John~Mark Kreikebaum, Vladislav~D. Kurilovich, Nathan Lacroix, David
  Landhuis, Tiano Lange-Dei, Brandon~W. Langley, Pavel Laptev, Kim-Ming Lau, Loïck~Le Guevel, Justin Ledford, Kenny Lee, Yuri~D. Lensky, Shannon Leon, Brian~J. Lester, Wing~Yan Li, Yin Li, Alexander~T. Lill, Wayne Liu, William~P. Livingston, Aditya Locharla, Erik Lucero, Daniel Lundahl, Aaron Lunt, Sid Madhuk, Fionn~D. Malone, Ashley Maloney, Salvatore Mandrá, Leigh~S. Martin, Steven Martin, Orion Martin, Cameron Maxfield, Jarrod~R. McClean, Matt McEwen, Seneca Meeks, Anthony Megrant, Xiao Mi, Kevin~C. Miao, Amanda Mieszala, Reza Molavi, Sebastian Molina, Shirin Montazeri, Alexis Morvan, Ramis Movassagh, Wojciech Mruczkiewicz, Ofer Naaman, Matthew Neeley, Charles Neill, Ani Nersisyan, Hartmut Neven, Michael Newman, Jiun~How Ng, Anthony Nguyen, Murray Nguyen, Chia-Hung Ni, Thomas~E. O'Brien, William~D. Oliver, Alex Opremcak, Kristoffer Ottosson, Andre Petukhov, Alex Pizzuto, John Platt, Rebecca Potter, Orion Pritchard, Leonid~P. Pryadko, Chris Quintana, Ganesh Ramachandran, Matthew~J. Reagor, David~M.
  Rhodes, Gabrielle Roberts, Eliott Rosenberg, Emma Rosenfeld, Pedram Roushan, Nicholas~C. Rubin, Negar Saei, Daniel Sank, Kannan Sankaragomathi, Kevin~J. Satzinger, Henry~F. Schurkus, Christopher Schuster, Andrew~W. Senior, Michael~J. Shearn, Aaron Shorter, Noah Shutty, Vladimir Shvarts, Shraddha Singh, Volodymyr Sivak, Jindra Skruzny, Spencer Small, Vadim Smelyanskiy, W.~Clarke Smith, Rolando~D. Somma, Sofia Springer, George Sterling, Doug Strain, Jordan Suchard, Aaron Szasz, Alex Sztein, Douglas Thor, Alfredo Torres, M.~Mert Torunbalci, Abeer Vaishnav, Justin Vargas, Sergey Vdovichev, Guifre Vidal, Benjamin Villalonga, Catherine~Vollgraff Heidweiller, Steven Waltman, Shannon~X. Wang, Brayden Ware, Kate Weber, Theodore White, Kristi Wong, Bryan W.~K. Woo, Cheng Xing, Z.~Jamie Yao, Ping Yeh, Bicheng Ying, Juhwan Yoo, Noureldin Yosri, Grayson Young, Adam Zalcman, Yaxing Zhang, Ningfeng Zhu, and Nicholas Zobrist.
\newblock Quantum error correction below the surface code threshold, 2024.

\bibitem{Grant_2018}
Edward Grant, Marcello Benedetti, Shuxiang Cao, Andrew Hallam, Joshua Lockhart, Vid Stojevic, Andrew~G. Green, and Simone Severini.
\newblock Hierarchical quantum classifiers.
\newblock {\em npj Quantum Information}, 4(1), December 2018.

\bibitem{farhi2018classificationquantumneuralnetworks}
Edward Farhi and Hartmut Neven.
\newblock Classification with quantum neural networks on near term processors, 2018.

\bibitem{Havl_ek_2019}
Vojtěch Havlíček, Antonio~D. Córcoles, Kristan Temme, Aram~W. Harrow, Abhinav Kandala, Jerry~M. Chow, and Jay~M. Gambetta.
\newblock Supervised learning with quantum-enhanced feature spaces.
\newblock {\em Nature}, 567(7747):209–212, March 2019.

\bibitem{kerenidis2019quantumalgorithmsdeepconvolutional}
Iordanis Kerenidis, Jonas Landman, and Anupam Prakash.
\newblock Quantum algorithms for deep convolutional neural networks, 2019.

\bibitem{Cong_2019}
Iris Cong, Soonwon Choi, and Mikhail~D. Lukin.
\newblock Quantum convolutional neural networks.
\newblock {\em Nature Physics}, 15(12):1273–1278, August 2019.

\bibitem{allcock2019quantumalgorithmsfeedforwardneural}
Jonathan Allcock, Chang-Yu Hsieh, Iordanis Kerenidis, and Shengyu Zhang.
\newblock Quantum algorithms for feedforward neural networks, 2019.

\bibitem{hernández2020imageclassificationquantummachine}
Héctor Iván~García Hernández, Raymundo~Torres Ruiz, and Guo-Hua Sun.
\newblock Image classification via quantum machine learning, 2020.

\bibitem{kiani2020quantummedicalimagingalgorithms}
Bobak~Toussi Kiani, Agnes Villanyi, and Seth Lloyd.
\newblock Quantum medical imaging algorithms, 2020.

\bibitem{Nakaji_2021}
Kouhei Nakaji and Naoki Yamamoto.
\newblock Quantum semi-supervised generative adversarial network for enhanced data classification.
\newblock {\em Scientific Reports}, 11(1), October 2021.

\bibitem{schuld2021effect}
Maria Schuld, Ryan Sweke, and Johannes~Jakob Meyer.
\newblock Effect of data encoding on the expressive power of variational quantum-machine-learning models.
\newblock {\em Physical Review A}, 103(3):032430, 2021.

\bibitem{schuld2021quantum}
Maria Schuld, Francesco Petruccione, Maria Schuld, and Francesco Petruccione.
\newblock Quantum models as kernel methods.
\newblock {\em Machine Learning with Quantum Computers}, pages 217--245, 2021.

\bibitem{kollias2023quantum}
Georgios Kollias, Vassilis Kalantzis, Theodoros Salonidis, and Shashanka Ubaru.
\newblock Quantum graph transformers.
\newblock In {\em ICASSP 2023-2023 IEEE International Conference on Acoustics, Speech and Signal Processing (ICASSP)}, pages 1--5. IEEE, 2023.

\bibitem{cherrat2024quantum}
El~Amine Cherrat, Iordanis Kerenidis, Natansh Mathur, Jonas Landman, Martin Strahm, and Yun~Yvonna Li.
\newblock Quantum vision transformers.
\newblock {\em Quantum}, 8(arXiv: 2209.08167):1265, 2024.

\bibitem{fan2023hybrid}
Fan Fan, Yilei Shi, Tobias Guggemos, and Xiao~Xiang Zhu.
\newblock Hybrid quantum-classical convolutional neural network model for image classification.
\newblock {\em IEEE transactions on neural networks and learning systems}, 2023.

\bibitem{wang2024shallow}
Aijuan Wang, Jianglong Hu, Shiyue Zhang, and Lusi Li.
\newblock Shallow hybrid quantum-classical convolutional neural network model for image classification.
\newblock {\em Quantum Information Processing}, 23(1):17, 2024.

\bibitem{Benedetti_2019}
Marcello Benedetti, Erika Lloyd, Stefan Sack, and Mattia Fiorentini.
\newblock Parameterized quantum circuits as machine learning models.
\newblock {\em Quantum Science and Technology}, 4(4):043001, November 2019.

\bibitem{mitarai2018quantum}
Kosuke Mitarai, Makoto Negoro, Masahiro Kitagawa, and Keisuke Fujii.
\newblock Quantum circuit learning.
\newblock {\em Physical Review A}, 98(3):032309, 2018.

\bibitem{Adhikary_2020}
Soumik Adhikary, Siddharth Dangwal, and Debanjan Bhowmik.
\newblock Supervised learning with a quantum classifier using multi-level systems.
\newblock {\em Quantum Information Processing}, 19(3), January 2020.

\bibitem{Schuld_2020}
Maria Schuld, Alex Bocharov, Krysta~M. Svore, and Nathan Wiebe.
\newblock Circuit-centric quantum classifiers.
\newblock {\em Physical Review A}, 101(3), March 2020.

\bibitem{acar2021covid}
Erdi Acar and Ihsan Yilmaz.
\newblock Covid-19 detection on ibm quantum computer with classical-quantum transfer learning.
\newblock {\em Turkish Journal of Electrical Engineering and Computer Sciences}, 29(1):46--61, 2021.

\bibitem{moradi2022clinical}
Sasan Moradi, Christoph Brandner, Clemens Spielvogel, Denis Krajnc, Stefan Hillmich, Robert Wille, Wolfgang Drexler, and Laszlo Papp.
\newblock Clinical data classification with noisy intermediate scale quantum computers.
\newblock {\em Scientific reports}, 12(1):1851, 2022.

\bibitem{houssein2022hybrid}
Essam~H Houssein, Zainab Abohashima, Mohamed Elhoseny, and Waleed~M Mohamed.
\newblock Hybrid quantum-classical convolutional neural network model for covid-19 prediction using chest x-ray images.
\newblock {\em Journal of Computational Design and Engineering}, 9(2):343--363, 2022.

\bibitem{toledo2022grading}
Santiago Toledo-Cort{\'e}s, Diego~H Useche, Henning M{\"u}ller, and Fabio~A Gonz{\'a}lez.
\newblock Grading diabetic retinopathy and prostate cancer diagnostic images with deep quantum ordinal regression.
\newblock {\em Computers in biology and medicine}, 145:105472, 2022.

\bibitem{azevedo2022quantum}
Vanda Azevedo, Carla Silva, and In{\^e}s Dutra.
\newblock Quantum transfer learning for breast cancer detection.
\newblock {\em Quantum Machine Intelligence}, 4(1):5, 2022.

\bibitem{parisi2022quantum}
Luca Parisi, Daniel Neagu, Renfei Ma, and Felician Campean.
\newblock Quantum relu activation for convolutional neural networks to improve diagnosis of parkinson’s disease and covid-19.
\newblock {\em Expert Systems with Applications}, 187:115892, 2022.

\bibitem{landman2022quantum}
Jonas Landman, Natansh Mathur, Yun~Yvonna Li, Martin Strahm, Skander Kazdaghli, Anupam Prakash, and Iordanis Kerenidis.
\newblock Quantum methods for neural networks and application to medical image classification.
\newblock {\em Quantum}, 6:881, 2022.

\bibitem{anagolum2024elivagarefficientquantumcircuit}
Sashwat Anagolum, Narges Alavisamani, Poulami Das, Moinuddin Qureshi, Eric Kessler, and Yunong Shi.
\newblock \'eliv\'agar: Efficient quantum circuit search for classification, 2024.

\bibitem{hanruiwang2022quantumnas}
Hanrui Wang, Yongshan Ding, Jiaqi Gu, Zirui Li, Yujun Lin, David~Z Pan, Frederic~T Chong, and Song Han.
\newblock Quantumnas: Noise-adaptive search for robust quantum circuits.
\newblock In {\em The 28th IEEE International Symposium on High-Performance Computer Architecture (HPCA-28)}, 2022.

\bibitem{Nation_2021}
Paul~D. Nation, Hwajung Kang, Neereja Sundaresan, and Jay~M. Gambetta.
\newblock Scalable mitigation of measurement errors on quantum computers.
\newblock {\em PRX Quantum}, 2(4), November 2021.

\bibitem{scikit-learn}
{Scikit-learn Developers}.
\newblock {\em ROC AUC Score — Scikit-learn Documentation}, 2025.
\newblock Accessed: January 3, 2025.

\bibitem{he2016deep}
Kaiming He, Xiangyu Zhang, Shaoqing Ren, and Jian Sun.
\newblock Deep residual learning for image recognition.
\newblock In {\em Proceedings of the IEEE conference on computer vision and pattern recognition}, pages 770--778, 2016.

\bibitem{Feurer2019}
Matthias Feurer, Aaron Klein, Katharina Eggensperger, Jost~Tobias Springenberg, Manuel Blum, and Frank Hutter.
\newblock {\em Auto-sklearn: Efficient and Robust Automated Machine Learning}, pages 113--134.
\newblock Springer International Publishing, Cham, 2019.

\bibitem{jin2019auto}
Haifeng Jin, Qingquan Song, and Xia Hu.
\newblock Auto-keras: An efficient neural architecture search system.
\newblock In {\em Proceedings of the 25th ACM SIGKDD international conference on knowledge discovery \& data mining}, pages 1946--1956, 2019.

\bibitem{google_automl_vision}
Google automl vision.
\newblock \url{https://cloud.google.com/vision/automl/docs}.

\end{thebibliography}

\end{document}